\documentclass[fleqn]{2023SCGE}
\usepackage{xcolor}
\usepackage{color}
\usepackage{float}
\usepackage{rotfloat}
\usepackage{rotating}
\usepackage{epsfig}
\usepackage{booktabs}
\usepackage{amsmath}
\usepackage{graphicx}
\usepackage{natbib,times}
\usepackage{amsmath}
\usepackage{url}
\usepackage{hyperref}
\usepackage{pdflscape}
\usepackage{ulem}
\usepackage{comment}

\usepackage{colortbl}
\usepackage{array}

\definecolor{dkblue}{RGB}{54, 86, 169}

\setcitestyle{numbers,square}

\newcommand{\mnras}{MNRAS}
\newcommand{\apj}{ApJ}
\newcommand{\apjs}{ApJS}
\newcommand{\apjl}{ApJL}

\newcommand{\aap}{A\&A}
\newcommand{\aaps}{A\&AS}
\newcommand{\apss}{Ap\&SS}
\newcommand{\nat}{Nature}

\newcommand{\araa}{ARAA}

\newcommand{\nar}{New Astron. Rev.}

\newcommand{\pasj}{PASJ}

\begin{document}

\ensubject{subject}

\ArticleType{Article}
\Year{2025}
\Month{??}
\Vol{??}
\No{??}
\DOI{10.1007/??}
\ArtNo{??}
\ReceiveDate{?? ??, 2025}
\AcceptDate{?? ??, 2025}

\title{A statistical study of type II outbursts of XRPs: Brighter accreting pulsars rotate faster}

\author[1,6]{Shan-Shan Weng}{wengss@njnu.edu.cn}

\author[2,3]{Xiao-Tian~Xu}{ xxu.astro@outlook.com}

\author[1]{Han-Long~Peng}{}

\author[1]{Yu-Jing~Xu}{}

\author[1]{Yan~Zhang}{}

\author[2,4]{\\Ying-Han~Mao}{}

\author[2,4]{Xiang-Dong~Li}{lixd@nju.edu.cn}

\author[5]{Jing-Zhi~Yan}{}

\author[5]{Qing-Zhong~Liu}{qzliu@pmo.ac.cn}

\AuthorMark{Weng S S}

\AuthorCitation{S.-S. Weng, X.-T. Xu, H.-L. Peng, Y.-J. Xu, Y. Zhang, Y.-H. Mao, X.-D. Li, J.-Z. Yan, and Q.-Z. Liu}
\footnotetext[1]{$*$ Corresponding author(s). E-mail(s): wengss@njnu.edu.cn; xxu.astro@outlook.com; lixd@nju.edu.cn; 
 qzliu@pmo.ac.cn.}

\address[1]{School of Physics and Technology, Nanjing Normal University, Nanjing, 210023, Jiangsu, China}
\address[2]{School of Astronomy and Space Science, Nanjing University, Nanjing, 210023, Jiangsu, China}
\address[3]{Argelander-Institut f\"ur Astronomie, Universit\"at Bonn, Auf dem H\"ugel 71, 53121 Bonn, Germany}
\address[4]{Key Laboratory of Modern Astronomy and Astrophysics, Nanjing University, Nanjing, 210023, Jiangsu, China}
\address[5]{Purple Mountain Observatory, Chinese Academy of Sciences, Nanjing 210008, China}
\address[6]{Nanjing key laboratory of particle physics and astrophysics, Nanjing, 210023, China}

\abstract{X-ray pulsars (XRPs) consist of a magnetized neutron star (NS) and an optical donor star. The NS accretes matter from the donor star producing pulsed X-ray emission. In most cases the donor stars are Be stars, and accretion is episodic, that is, the NSs are generally X-ray dim, but occasionally experience outbursts.  Here, we carry out a statistical study with the X-ray monitoring data, and obtain strong correlations between the spin periods of the NSs and the outburst parameters for the first time. We show that XRPs containing faster rotating NSs tend to display more violent eruptions. In addition, pulsating ultraluminous X-ray sources  in nearby galaxies follow the similar relationship. We demonstrate  that most of these systems are close to the spin equilibrium, and that brighter pulsars have acquired more angular momentum by accreting matter from their companion stars, resulting in faster rotating NSs.}



\maketitle

\begin{multicols}{2}
	
\section{Introduction}\label{sect:intro}

The scenario of accreting X-ray pulsars (XRPs) has been proposed to interpret X-ray pulsations detected from X-ray binaries since 1970s. The cyclotron resonant scattering features (CRSFs) emerging from the X-ray spectra confirm that the neutron stars (NSs) in XRPs are highly magnetized with a surface magnetic field of $B \sim 10^{11-13}$ G \citep{Staubert2019, Kong2022}.  Until now, more than 200 XRPs have been discovered in the Galaxy, Large and Small Magellanic Clouds (LMC and SMC) \citep{Liu2005, Liu2006, Fortin2023, Neumann2023} {\protect\url{http://www.iasfbo.inaf.it/\~mauro/pulsar\_list.html}}. While part of XRPs are persistent sources with a luminosity of $L_{\rm X} \sim 10^{34-38}$ erg/s, the majority are transient, and the complex interaction between the magnetosphere of NSs and their surrounding material results in dramatic observational phenomena \citep[for a review see e.g. ][]{Reig2011, Weng2024}. 

A large number of transient XRPs are revealed in Be X-ray binaries (BeXRPs), in which the Be stars supply the accreting material via both a fast polar wind and a slow equatorial outflow, displaying two types of outbursts --- the so-called type I and type II outbursts. It has been already clear that, type I outbursts are quasi-periodic and are triggered as the NSs capture more gas at the periastron passage. These events are short-lived, lasting a relatively small fraction of the orbital period ($< 0.3~P_{\rm orb}$), and their peak luminosities generally do not exceed $10^{37}$ erg/s \citep{Okazaki2001, Okazaki2013}. 

In comparison, type II (or giant) outbursts are much less frequent and significantly brighter. The outburst events are not locked to the orbital phase and the typical timescale can be as long as  several months.  The apparent X-ray luminosities at the peak of the outbursts approach or exceed the Eddington luminosity ($L_{\rm peak} \sim 10^{37-39}$ erg/s), making them among the brightest objects in the X-ray sky.  During outbursts, the pulsars in BeXRPs rapidly spin up, and the X-ray pulse profiles switch between single-peak and double-peaks, suggesting a transition of the accretion pattern. In addition to the large spin-up rates,  low frequency ($\sim 10-100$ mHz) quasi-periodic oscillations are detected, indicating that transient accretion disks are temporarily formed around the NSs during outbursts \citep{Finger1996b, Bozzo2009}.  

It is recognized that, as the mass accretion rate increases, XRPs go through three distinguished states: ejector, propeller and accretor, depending on interaction between the magnetosphere and the material supplied by the donor star \citep{Davies1981, Ikhsanov2001, Shakura2012}. The size of the magnetosphere is determined by the balance between the ram pressure of the accreting flow and the magnetic pressure. 

{\it The ejector regime.} If the rate of infalling gas is extreme low, $R_{\rm mag}$ is larger than the light cylinder radius, $R_{\rm LC} = \frac{cP_{\rm spin}}{2\pi}$, the NS is powered by the rotational energy and behaves as a pulsar. That is, the spin period, $P_{\rm spin}$ increases with time during the quiescence state.

{\it The propeller regime.} When the rate of infalling matter increases, the ram pressure overcomes the NS's radiative pressure, and the accreting gas penetrates inside the light cylinder and falls down towards the NS magnetosphere. But accretion is still inhibited by the centrifugal barrier of the NS, because $R_{\rm mag}$ exceeds the corotation radius, $R_{\rm cor} = (\frac{GMP_{\rm spin}^{2}}{4\pi^{2}})^{1/3}$. The infalling gas is accelerated outward and takes away the angular momentum of NSs, leading to the spin-down of pulsars. This regime ends when $R_{\rm mag} = R_{\rm cor}$, which also determines the equilibrium period, $P_{\rm eq} \propto B^{6/7}~\dot{M}^{-3/7}$, where $B$ is the surface magnetic field of the NS and $\dot{M}$ is the accretion rate.

{\it The accretion regime.} At even higher accretion rate,  $R_{\rm mag}$ becomes smaller than $R_{\rm cor}$ (or $P_{\rm spin} > P_{\rm eq}$), the matter can be accreted on to the surface of NSs.  In the meantime,  NSs are spun up due to the angular momentum transferred from the accretion flow.

The spin evolution due to the angular momentum exchanging between accreting flows and pulsar, provides a crucial diagnostic for estimating magnetic filed of XRPs \citep[e.g. ][]{Weng2017, Chashkina2019, Liu2022}. However, several key issues remain controversial. For instance, after long-term evolution, do the spin periods of BeXRPs approach equilibrium periods? Various torque models, involving disk and wind accretion, and being close to spin equilibrium or not, have been proposed \citep[e.g. ][]{Klus2014, Xu2019}.
 
In this work, we perform a statistical study on the giant outbursts of BeXRPs for the first time. The manuscript is formatted as follows. In Section~\ref{sec:data_method}, we collect the sample of giant outbursts, and analyze their light curves. In Section~\ref{sec:results}, we investigate the correlations among the outburst parameters and the binaries' parameters (e.g. orbital and spin periods). In Section~\ref{sec:discussion}, we interpret the strong anti-correlation between the X-ray luminosity and the spin period with the spin evolution model. Summary is made in Section~\ref{sec:sum}.

\section{Data analysis}\label{sec:data_method}

\subsection{Sample}\label{sec:sample}

In the last half-century, dozens of X-ray missions were launched, and conducted frequent and thorough inspections of XRPs. When accreting pulsars enter into giant outbursts, they become the brightest X-ray sources in the sky. Detailed spectral and temporal properties of outbursts can be extracted from pointed observations using multiple X-ray telescopes, e.g. {\it IXPE}, {\it INTEGRAL}, {\it RXTE}, {\it Swift}, {\it XMM-Newton}, {\it Chandra}, {\it NuStar}, {\it Insight-HXMT}, {\it AstroSat}, and  {\it NICER}. However, the pointed observations are usually sparsely sampled, and only cover part of outbursts. Alternatively, the monitoring data from telescopes having large field-of-view trace the whole outburst, and are suitable for a statistical study. {\it RXTE} All Sky Monitor (ASM) and {\it Swift} Burst Alert Telescope (BAT) are two highly sensitive, large field-of-view instruments designed to monitor X-ray transients. Their data provide a baseline of around 30 years to study the statistical properties of transient XRPs. In this work, we use the daily light curves made by {\it RXTE}/ASM and {\it Swift}/BAT. 

{\it RXTE} was launched on December 30, 1995, and  decommissioned on January 5, 2012. It carries two pointed instruments, the Proportional Counter Array, the High Energy X-ray Timing Experiment, and an All-Sky Monitor (ASM),  which scans $\sim 80\%$ of the sky in the energy band of 2--12 keV every orbit. More than 26 giant outbursts had been monitored by ASM, and 12 of them occurred before the operation of {\it Swift}.  The ASM daily light curves can be downloaded  from the ASM webpage: {\protect\url{https://xte.mit.edu/ASM_lc.html}}. It worth to note that, XRPs are hard X-ray transients, i.e. they emit most of their radiation in the hard X-ray band. Besides, the {\it Swift}/BAT is more sensitive than the {\it RXTE}/ASM. Therefore, for the outbursts occurred after 2004, we only use the {\it Swift}/BAT data. 

{\it Swift} was designed to study Gamma-ray bursts (GRBs) in broadband with its three payloads: the BAT, the X-ray Telescope, and the UV/optical Telescope. BAT has a large field-of-view and high sensitivity, and therefore is suitable to detect not only GRBs but also some other interesting transients.  Since {\it Swift} was launched on November 20, 2004, BAT has completely monitored more than 38 giant outbursts from XRPs. The 15-50 keV BAT daily light curves can be retrieved from the public monitor web page: {\protect\url{https://swift.gsfc.nasa.gov/results/transients/}} \citep{Krimm2013}. In some cases, because the monitoring was interrupted due to the solar aspect angle violation or other instrument issues, the outburst profile was unavailable and we discard the outburst, e.g. the 2005 giant outburst of 1A~0535+262. There are several anomalous data points with ultra-high count rate shown in the daily light curve of the 2023 giant outburst of 4U~0115+634. We check the  {\it NICER} and the {\it MAXI} light curves, but do not find any flux jump during this activity. Because the real peak flux of 2023 outburst is lower than the value recorded in the 1999 outburst, the 2023 giant outburst of 4U~0115+634 is excluded in this work.

RX~J0209.6-7427 is located in the outer wing of SMC, and the rising part of its 2019 super-Eddington outburst was missed by the {\it Swift}/BAT. Alternatively, {\it NICER} carried out the high-cadence observations covering the whole outburst\cite{Vasilopoulos2020}.  Therefore, we extract the 2--12 keV light curve from all 92 {\it NICER} observations for the outburst parameters estimation. The data are filtered with the standard criteria using  \textsc{heasoft}  (version 6.30), and the count rates are averaged for each individual observation. 

The peculiar low-mass X-ray binary, GRO~J1744-28 possesses a strong magnetic field of ($\sim 5\times10^{11}$ G) and a relatively slow spin period 467 ms \citep{DAi2015}, distinguishing it from typical LMXBs -- such as accreting millisecond X-ray pulsars or non-pulsating neutron stars with weak magnetic fields ($\sim 10^{8}$ G). Alternatively, its X-ray properties resemble those of young accreting pulsars. Therefore, we also include GRO~J1744-28 in this work.

In order to avoid possible anomalous fluctuations,  only the data points with $> 5\sigma$ significance  are used to measure the outburst parameters. We further discard the outbursts having sparse data, i.e. less than 10 data points. Eventually, we collect the light curves of 51 giant outbursts from 23 BeXRPs and a low-mass X-ray binary, GRO~J1744-28, which were completely recorded, 12 from {\it RXTE}/ASM, 38 from {\it Swift}/BAT, and 1 from {\it NICER}. The main parameters of these XRPs are summarized in Table \ref{tab:log}.

\begin{table*}
    \resizebox{1.0\textwidth}{!}{
    \begin{tabular}{ccccccccc}
    \hline
     Source & $d$ (kpc) &$P_{\rm spin}$ (s)&$P_{\rm orb}$ (days)& $e$ & $E_{\rm cyc}$ (keV) & $B$ (10$^{12}\,$G) &Epoch of Outburst& Ref. \\
     \hline
SMC X-3 &  $62.1\pm1.9$ &   7.8 &   44.5 & 0.26 & -- & -- &{\it 2016} & \cite{Weng2017}  \\     
4U~0115+63  & $5.8_{-0.4}^{+0.8}$ & 3.6 & 24.3 & 0.34 & 12 & 1.2 & {\it 1999}; 2000; 2004; 2008;  & \cite{Okazaki2001, Staubert2019}\\   
& & &  & & & & 2011; 2015; 2017 & \\
RX~J0209.6-7427 & $49.6\pm0.6$ &  9.3 & 48.0 & 0.317 & -- & -- &{\it 2019}$^{\dag}$ & \cite{Chandra2020, Hou2022} \\ 
Swift~J0243.6+6124 & $5.2_{-0.3}^{+0.3}$ & 9.86 & 27.6 & 0.098 & 133 &13.8 & {\it 2017} & \cite{Wilson2018, Kong2022}\\
V~0332+53 & $5.6_{-0.5}^{+0.7}$ & 4.37 & 33.85 & 0.371 & 28 & 2.9 & {\it 2004}; 2015 & \cite{Doroshenko2016, Staubert2019}\\
LS~V~+44~17 & $2.4_{-0.1}^{+0.1}$ & 202.5 & -- & -- & -- & -- & {\it 2023} & \cite{Reig1999}\\
RX~J0520.5-6932 & $49.6\pm0.6$ &  8.0 & 23.9 & 0.03 & -- & -- &{\it 2014} & \cite{Kuehnel2014} \\    
1A~0535+262 & $1.79_{-0.07}^{+0.08}$ & 104 & 110.3 & 0.47 & 50 & 5.2 & 2009; 2011; {\it 2020} & \cite{Finger1996b, Staubert2019}\\
MXB~0656-072 & $5.7_{-0.5}^{+0.5}$ & 160.4 & 101.2 & 0.4 & 33 & 3.4 &{\it 2003} & \cite{Yan2012, Staubert2019} \\
GRO~J1008-57 & $3.5_{-0.1}^{+0.2}$ & 93.5 & 249.5 & 0.68 & 90 & 9.3 &{\it 2012}; 2015; 2017; 2020  & \cite{Riquelme2012, Kuehnel2012, Ge2020}\\
1A~1118-616 & $2.9_{-0.1}^{+0.1}$ & 407.7 & 24.0 & 0.10 & 55 & 5.7 &{\it 2009} & \cite{Staubert2011, Staubert2019}\\   
MAXI~J1409-619 &  $4.4_{-0.6}^{+1.2}$ &  504 & 14.7 & -- & 44 & 3.7 & {\it 2010} & \cite{Orlandini2012, Donmez2020}\\
2S~1417-624 & $7.4_{-1.8}^{+3.1}$ & 17.5 &  42.1 & 0.45 & -- & -- &2009; {\it 2018} & \cite{Finger1996a} \\   
GRO~J1750-27 & $18.0_{-4.0}^{+4.0}$ & 4.45 & 29.8 & 0.36 & 44 & 4.6 &{\it 2008}; 2015; 2021 & \cite{Lutovinov2019, Malacaria2020, Malacaria2023} \\
XTE~J1858+034 & $10.9_{-1.0}^{+1.0}$ & 218.4 & -- & -- & 48 & 5.0 &{\it 1998}; 2019 & \cite{Malacaria2021}\\
XTE~J1859+083 & $8.7_{-5.1}^{+3.6}$ & 9.8 & 38.0 & -- & -- & --&{\it 2015} & \cite{Malacaria2020, Salganik2022}\\
4U~1901+03 & $12.4_{-0.2}^{+0.2}$ & 2.76 & 22.5 & 0.014 & 30? & 3.1? & {\it 2003}; 2019 & \cite{Tuo2020, Beri2021}\\   
XTE~J1946+274 & $13.1_{-2.3}^{+3.2}$ & 15.8 & 172.7 & 0.246 & 38 & 3.9 &{\it 2010} & \cite{Marcu2015, Doroshenko2017}\\  
KS~1947+300 & $15.1_{-2.6}^{+3.2}$ & 18.8 & 40.4 & 0.033 & 12? & 1.2? &{\it 2000}; 2013 & \cite{Galloway2004, Furst2014} \\   
EXO~2030+375 & $2.4_{-0.4}^{+0.5}$ & 41.3 & 46.0 & 0.41 & -- & -- &{\it 2006}; 2021 & \cite{Wilson2008, Yang2024}\\
GRO~J2058+42 & $8.9_{-1.0}^{+1.0}$ & 195 & 110 & -- & 10 & 1.0 &{\it 2019} & \cite{Wilson1998, Molkov2019} \\ 
SAX~J2103.5+4545 & $6.2_{-0.4}^{+0.6}$ & 358.6 & 12.7 & 0.40 & -- & -- & {\it 2007}; 2010; 2016 & \cite{Camero2007} \\  
Cep~X-4 & $7.4_{-0.5}^{+0.6}$ & 66.3 & -- & -- & 30 & 2.6 & {\it 1997}; 2002;  2009; 2014; & \cite{Wilson1999, Staubert2019}\\
 & & & & & & &  2018 & \\
\hline
GRO~J1744-28 & 8.0 & 0.467 & 11.83 & 0.0 & 4.7? & 0.5? & {\it 1995}; 1997; 2014  & \cite{Bildsten1997, DAi2015, Sanna2017}\\
\hline
    \end{tabular}
    }
    \caption{Main system properties of XRPs, which are listed in order of right ascension. 
    $E_{\rm cyc}$: The centroid line energies of CRSF, and their errors are about 10\%. The corresponding magnetic fields $B$ are estimated by $E_{\rm cyc}=11.6\,\text{keV}\,(B/10^{12}\,\text{G})\, (1+z)^{-1}$ (Eq. 1 in Ref.\citep{Staubert2019}), where gravitational redshift $z$ is taken to be 0.2 in our estimation. While the CRSF measurement of KS\,1947+300 with {\it NuSTAR} may be affected by the presence of flares \citep{Manikantan2023}, it does not affect the main result of this work. 
    Epoch of Outburst: The brightest outburst for each source is shown with the italic font. $\dag$: Parameters are estimated with the {\it NICER} data.   }
    \label{tab:log}
\end{table*}

\subsection{X-ray light curves of XRPs}

For this work, the main source of the uncertainty is the distance. The distances of $49.6\pm0.6$ kpc\cite{Pietrzynski2019} and $62.1\pm1.9$ kpc \citep{Graczyk2014} are adopted for the LMC and the SMC, respectively. For the Galaxy sources, the X-ray luminosities are calculated with the distances obtained from the third {\it Gaia} Data Release (DR3). We retrieve the {\it Gaia} source ID for each source from the {\it Gaia} Archive ({\protect\url{https://gea.esac.esa.int/archive/}}), and then query the distance of the correspondent {\it Gaia} source ID  from the {\it Gaia} DR3 ({\protect\url{https://vizier.cds.unistra.fr/viz-bin/VizieR?-source=I/352}}). However, because of the heavy absorption and/or too far away ($>$ 10 kpc), the {\it Gaia} distances are unavailable for GRO~J1750-27,  XTE~J1858+034, 4U~1901+03, and GRO~J1744-28.  In this case, we opt for the distances reported in literature, which were estimated with other methods instead of parallaxes (Table \ref{tab:gaia}), e.g. based on the evolution of the pulse profiles \citep{Tuo2020} or the spin-up measurement \citep{Malacaria2021}.

The X-ray photon count rates are converted into the unit of Crab, with 1 Crab = 0.22 count~s$^{-1}$~cm$^{-2}$ for the {\it Swift}/BAT,  75 count~s$^{-1}$ for {\it RXTE}/ASM, and 2537 count~s$^{-1}$ for {\it NICER}/XTI in 2--12 keV, respectively. The X-ray luminosity is further calculated by adopting that the 1 Crab = $2\times10^{-8}$ erg~cm$^{-2}$~s$^{-1}$. Four parameters are defined to characterize the outburst properties: (1) the X-ray peak luminosity, $L_{\rm peak}$, (2) the duration of the outburst $T_{\rm 50}$ defined as the time with the X-ray luminosity being larger than $0.5 \times L_{\rm peak}$, estimated with the linear interpolation method, (3) the total energy radiated during $T_{\rm 50}$, $E_{\rm tot}$, and (4) the mean X-ray luminosity during $T_{\rm 50}$, $\bar{L}_{\rm X} = E_{\rm tot}/T_{\rm 50}$. The $1\sigma$ errors of the outburst parameters ($T_{\rm 50}$,  $E_{\rm tot}$, and $\bar{L}_{\rm X}$) are estimated with the the count randomization method, wherein 2,000 realizations of light curves with count rate measurements are adjusted by random Gaussian deviates scaled to the measurement uncertainties. All the original data are shown in Figures \ref{fig:bright} and \ref{fig:faint}, but only the data with $> 5\sigma$ source detection are used to calculate the outburst parameters and their $1\sigma$ errors.

Since spectra of XRPs with a photon index of $\Gamma \sim 0.5-1$ are significantly harder than the spectrum of Crab ($\Gamma \sim 2.07$) \citep{Kirsch2005}, there are systematic differences in the fluxes and the light curve profiles derived in different energy band. Considering all the outbursts occurred during 2005 and 2011, the fluxes recorded by {\it Swift}/BAT are about 2--4 times of the values obtained from the {\it RXTE}/ASM data. Thus, the fluxes derived in the energy band of $2-12$ keV are then multiplied by the median ($\sim$3) before the statistical analysis. Meanwhile, the differences in the derived $T_{50}$ and the fluence (with the flux correction) are less than 20\%, which are much less than the individual differences. We remain the values of $T_{50}$ and the $E_{\rm tot}$ estimated from the original light curves.

\begin{table}[H]
    \centering
    \resizebox{0.45\textwidth}{!}{
    \begin{tabular}{ccc}
    \hline
     Source & {\it Gaia-}EDR3 Source ID & $d$ (kpc) \\
     \hline
4U~0115+63  & 524677469790488960 & $5.8_{-0.4}^{+0.8}$ \\
Swift~J0243.6+6124 & 465628193526364416 & $5.2_{-0.3}^{+0.3}$ \\
V~0332+53 & 444752973131169664 & $5.6_{-0.5}^{+0.7}$ \\
LS~V~+44~17 & 252878401557369088 & $2.4_{-0.1}^{+0.1}$ \\
1A~0535+262 & 3441207615229815040 & $1.79_{-0.07}^{+0.08}$ \\
MXB~0656-072 & 3052677318793446016 & $5.7_{-0.5}^{+0.5}$ \\
GRO~J1008-57 & 5258414192353423360 & $3.5_{-0.1}^{+0.2}$ \\
1A~1118-616 & 5336957010898124160 &  $2.9_{-0.1}^{+0.1}$ \\  
MAXI~J1409-619 &  5866114062855956224 & $4.4_{-0.6}^{+1.2}$ \\
2S~1417-624 & 5854175187710795136 & $7.4_{-1.8}^{+3.1}$  \\ 
GRO~J1750-27 &   		$^{*}$		   & $18.0_{-4.0}^{+4.0}$ \cite{Lutovinov2019} \\
XTE~J1858+034 &   		$^{*}$	   & $10.9_{-1.0}^{+1.0}$ \cite{Malacaria2021}\\     
XTE~J1859+083 & 4310118570535453696 & $8.7_{-5.1}^{+3.6}$\\  
4U~1901+03 &   		$^{*}$		   & $12.4_{-0.2}^{+0.2}$ \cite{Tuo2020}\\  
XTE~J1946+274 & 2028089540103670144 & $13.1_{-2.3}^{+3.2}$\\  
KS~1947+300 & 2031939548802102656 & $15.1_{-2.6}^{+3.2}$ \\ 
EXO~2030+375 & 2063791369815322752 & $2.4_{-0.4}^{+0.5}$ \\
GRO~J2058+42 & 2065653598916388352 & $8.9_{-1.0}^{+1.0}$ \\ 
SAX~J2103.5+4545 & 2162805896614571904 & $6.2_{-0.4}^{+0.6}$ \\
Cep~X-4 & 2178178409188167296 & $7.4_{-0.5}^{+0.6}$ \\
\hline
GRO~J1744-28 &   		$^{*}$		   & 8.0 \cite{Bildsten1997} \\
\hline
    \end{tabular}
    }
    \caption{Distances of XRPs in the Galaxy.  Sources are listed in order of right ascension. $^{*}$:  {\it Gaia} distance is unavailable for these sources, and we use the distance reported in literature.}
    \label{tab:gaia}
\end{table}


\section{Results}\label{sec:results}

\begin{figure*}
\centering
\includegraphics[width=0.9\textwidth]{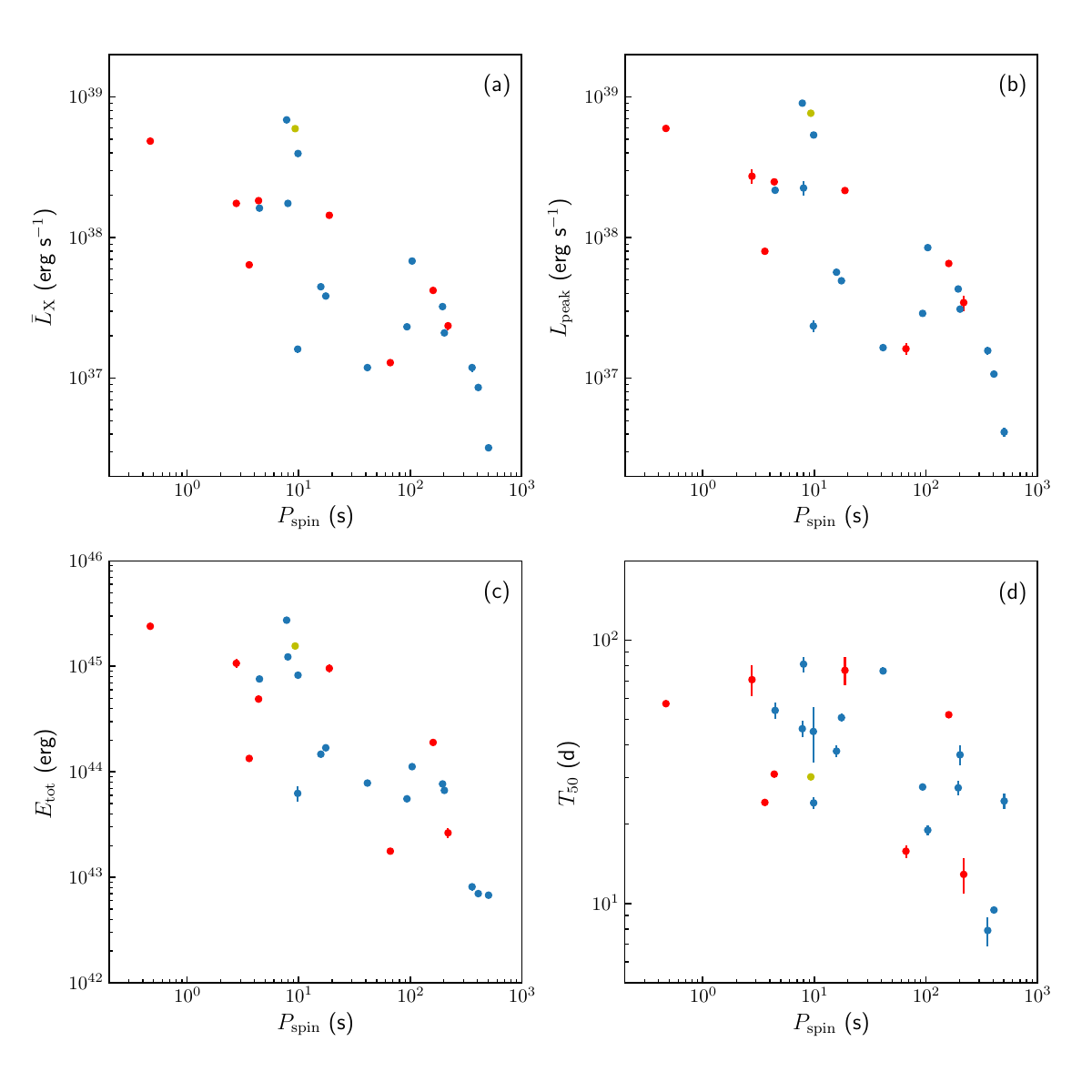}
\caption{Anti-correlations between $P_{\rm spin}$ and the outburst parameters. Blue, red, and green circles correspond to the {\it Swift}/BAT, {\it RXTE}/ASM, and {\it NICER} data, respectively.  It is worth to note that, the $1\sigma$ errors are plotted in the figure; however, they are smaller than the symbol sizes in most cases.
\label{fig:bright_corr}
}
\end{figure*}

The Spearman's rank correlation coefficient ($\rho$) and the corresponding significance level ($p$-value) are calculated to estimate the correlations between each pair of parameters (Table \ref{tab:corr}).  Apparently, the spin periods ($P_{\rm spin}$) have strong anti-correlations with all the outburst parameters, $L_{\rm peak}$, $T_{\rm 50}$,  $E_{\rm tot}$,  and $\bar{L}_{\rm X}$ (Figures \ref{fig:bright_corr} and \ref{fig:all_corr}), with $\rho < -0.547$ and $p < 0.006$. In contrast, there is no significant correlation between the outburst parameters and the orbital period $P_{\rm orb}$ and the eccentricity  $e$ (Table \ref{tab:corr}). 

To confirm these relationships, we carry out several additional tests taking into account the possible influence of the following factors. (1) The luminosity uncertainty. Because the spectral shapes of XRPs generally evolve with the X-ray luminosity, the precise bolometric corrections rely on the high energy resolution pointed observations, which are not always available. Alternatively, assuming different spectral profiles ($\Gamma \sim 0.5-1$ and $nH \sim 10^{21} - 5 \times 10^{22}$ cm$^{-2}$), we use the tool of  ``WebPIMMS"\footnote{{\protect\url{https://heasarc.gsfc.nasa.gov/cgi-bin/Tools/w3pimms/w3pimms.pl}}} to estimate that the correction factor. Its value varies by less than 30\% for {\it Swift}/BAT, but can change by up to a factor of 2-3  for {\it RXTE}/ASM. We suggest that the uncertainty of the outburst parameters derived without the bolometric correction (or with a constant correction factor) is much smaller than the individual differences.  We therefore do not try to make the bolometric correction. Nevertheless, all correlations can be found with the {\it Swift}/BAT outbursts alone (22 outbursts, $p < $ 0.004 and $\rho < -0.599$). (2) The distance uncertainty. The {\it Gaia} distances are unavailable for GRO~J1750-27,  XTE~J1858+034, 4U~1901+03, and GRO~J1744-28, and their distances might be quite different from the values used in this Paper. Excluding these four sources, we find that the anti-correlations between $P_{\rm spin}$ and the outburst parameters become slightly weaker, but still significant ($p < 2.3\times10^{-4}$ and $\rho < -0.747$), except for $T_{\rm 50}$ ($\rho/p = -0.448/0.049$). (3) Specific outbursts. In this sample, MAXI~J1409-619 is the slowest rotating XRP, and the peak luminosity during its 2010 outburst is only $\sim 4 \times 10^{36}$ erg/s. The duration of the outburst $T_{\rm 50} \sim 24.5\pm1.5$ days is much longer than its orbital period $P_{\rm orb} = 14.7$ days. Hence, this outburst should be considered as a giant outburst. Nevertheless, the strength of correlations do not change much when excluding the outburst of MAXI~J1409-619. Alternatively, the low-mass X-ray binary, GRO~J1744-28 experienced the most luminous giant outburst in 1995, which was brighter than the outbursts in 1997 and 2014 by a factor of 2-3. However, the outburst was not well sampled by the {\it RXTE}/ASM data (panel 24 of Figure \ref{fig:bright}), and the values of all outburst parameters were underestimated. Meanwhile, increasing the values of outburst parameters by tens of percent (or more) should further improve the anti-correlations. (4) Extended sample. We also take all 51 giant outbursts into account, treat them as independent events, and confirm all these anti-correlations ($\rho/p = -0.512/1.2\times10^{-4}$ for $P_{\rm spin}$ v.s. $T_{50}$). 


\begin{table*}[!t]
\begin{center}
    \resizebox{0.8\textwidth}{20mm}{
    \begin{tabular}{|c|c|c|c|c|c|c|c|c|}
    \hline
    \rowcolor{brown!30} \cellcolor{white} $\rho/p$ & $L_{\rm X}$ & $E_{\rm tot}$ & $T_{\rm 50}$ & $L_{\rm peak}$ & $P_{\rm spin}$ & $P_{\rm orb}$ & $e$  & $E_{\rm cyc}$\\
      \hline
       \rowcolor{blue!10} \cellcolor{brown!30}$L_{\rm X}$   & ---  &   0.935/2.2e-11   &   0.405/0.050   &    0.995/1.0e-23  & {\bf -0.775/8.8e-6}  & -0.038/0.871  &  -0.526/0.025 &  -0.190/0.449 \\
      \hline
       \rowcolor{blue!5} \cellcolor{brown!30} $E_{\rm tot}$   &  0.935/2.2e-11    &  ---    &  0.665/5.4e-4  &  0.950/1.9e-6 & {\bf -0.790/7.0e-6}   & -0.081/0.728  & 0.632/0.006 &  -0.296/0.234 \\
      \hline
       \rowcolor{blue!10} \cellcolor{brown!30} $T_{\rm 50}$   &  0.405/0.050   &    0.665/5.4e-4  & ---  &  0.442/0.032 & {\bf -0.547/6.4e-3}   & -0.017/0.944   & -0.385/0.116 & 0.388/0.112 \\     
       \hline
       \rowcolor{blue!5} \cellcolor{brown!30} $L_{\rm peak}$   &   0.995/1.0e-23   &  0.950/1.9e-6   &  0.442/0.032  & ---  &   {\bf -0.773/1.5e-5} & 0.042/0.859  & 0.534/0.024 &  -0.139/0.583 \\  
       \hline
      \rowcolor{blue!5} \cellcolor{brown!30}$P_{\rm spin}$ &   {\bf -0.775/8.8e-6}    &  {\bf -0.790/7.0e-6}    & {\bf -0.547/6.4e-3}   &  {\bf -0.773/1.5e-5} &  ---  & 0.253/0.267  & 0.514/0.031 &  0.198/0.432 \\
      \hline
       \rowcolor{blue!10} \cellcolor{brown!30} $P_{\rm orb}$   &  -0.038/0.871    &    -0.081/0.728 & -0.017/0.944  &  0.042/0.859  & 0.253/0.267  & ---   & 0.591/0.011 & 0.216/0.390 \\    
       \hline
       \rowcolor{blue!5}  \cellcolor{brown!30} $e$   &  -0.526/0.025     &   0.632/0.006   & -0.385/0.116   & 0.534/0.024  &  0.514/0.031 & 0.591/0.011  & ---  & 0.040/0.875 \\  
\hline
       \rowcolor{blue!10} \cellcolor{brown!30} $E_{\rm cyc}$   &  -0.190/0.449    &    -0.296/0.234 & 0.388/0.112  & -0.139/0.583  & 0.198/0.432  &  0.216/0.390   & 0.040/0.87 &  --- \\    
       \hline
    \end{tabular}
    }
    \caption{Spearman's rank correlation estimated with the data of the brightest outbursts for each source. }
    \label{tab:corr}
    \end{center}
\end{table*}



\section{Discussion}\label{sec:discussion}

When checking the XRPs catalogues, we could barely see the weak anti-correlation between the maximum luminosity and $P_{\rm spin}$, see e.g., \citep{Raguzova2005, Sidoli2018}.  All these catalogues consist of diverse populations of XRPs, which are fed by distinct accretion flows. For instance, in contrast to the transient XRPs discussed in this work, a small group of persistent XRPs containing slowly rotating NSs ($P_{\rm spin} > 200$ s) have lower luminosity ($L_{\rm X} < 10^{35}$ erg/s), and these systems are likely undergo spherical accretion. Only when taking the giant outbursts of XRPs into account, the correlations become remarkable. XRPs during giant outbursts are obviously far from the spin equilibrium related to the X-ray luminosity of giant outbursts, and are always spinning up, which might explain the derivation of the observed $\bar{L}_{\rm X}\text{--}P_{\rm spin}$ correlation from $L_{\rm X}\propto P_{\rm eq}^{3/7}$ expected from the static disk-accretion-induced spin evolution model. The typical timescale of the spin evolution of a disk accreting NS can be estimated as
\begin{equation}
    \tau_{\rm spin} \sim \left|\frac{P_{\rm spin}}{\dot{P}_{\rm spin}}\right| \sim \left(\frac{10^5}{4.3\mu_{30}^{2/7}P_{\rm spin}L_{37}^{6/7}}\right)\,\text{\rm yr},
\end{equation}
where the spin period change rate $\dot{P}_{\rm spin}$ is estimated by using Eq.\,(2) in Ref.\cite{Ho2014MNRAS.437.3664H}, $\mu_{30}$ is the magnetic moment of the NS in the unit of $10^{30}\,\text{G}\,\text{cm}^3$, and $L_{\rm 37}$ is the X-ray luminosity in the unit of $10^{37}\,\text{erg}\,\text{s}^{-1}$. With a typical outburst luminosity of giant outbursts, $10^{38}\,\text{erg}\,\text{s}^{-1}$, a magnetic field of $10^{12}\,$G, and a $P_{\rm spin}$ of 10\,s, we have  $\tau_{\rm spin}\sim10^3$\,yr, which is shorter with a stronger magnetic field. Since it is orders of magnitude shorter than the lifetime of BeXRBs\footnote{The lifetime of a BeXRB is comparable to the main-sequence lifetime of the Be stars. According to the stellar evolution model in Ref.\citep{Xu2025arXiv250323876X}, the lifetime $\tau_{\rm MS,Be}$ of a 10\,$M_\odot$ Be star is about 26\,Myr, which is much longer than $\tau_\text{eq}$. Assuming a constant star formation rate, the chance of detecting a system should be proportional to its lifetime (see appendix in Ref.\cite{Xu2025arXiv250323876X}). Hence, the maximal fraction of BeXRPs that are far away from their spin equilibrium should be on the order of $\tau_{\rm spin}/\tau_{\rm MS,Be}$, which is $\sim3\times10^{-5}$. }, we find it is unlikely that many BeXRPs are still far away from its spin equilibrium according to the traditional disk accretion theory, especially for the XRPs with relatively strong magnetic fields. The above considerations show that static accretion model have some difficulties in understanding the spin evolution in transient XRPs.

Therefore, in this study, we present an alternative way to understand the spin evolution of XRPs based on the spin evolution model established in Ref. \cite{Xu2019}, which is motivated by the observational facts that the vast majority of BeXRPs show strong X-ray variability. The authors propose that XRPs at different luminosity states have different accretion modes (see also Ref. \citep{Cheng2014} for observational evidence), which exert different torques on the NSs \citep{Xu2019} (hereafter multi-mode spin evolution model). Different from the static single-mode accretion model, this multi-mode model suggests that the occurrence of short-term spin-ups and spin-downs does not contradict a long-term spin equilibrium (see Section \ref{multi-mode-model}). One observational consequence of this multi-mode model is that the inferred magnetic field of a BeXRP by using one luminosity state should be lower than the magnetic field inferred by using the long-term averaged spin evolution over various luminosity states.
The reason is that the presence of the quiescent phase reduces the overall spin-up torque, causing an potential overestimation of magnetic field by using the canonical single-mode disk accretion model. Taking  SXP\,15.6 as an example, its spin evolution during an outburst leads to a magnetic field of about $5\times10^{11}\,$G \citep{Vasilopoulos2022}, while it is about $2\text{--}4\times10^{12}\,$G if it is close to the spin equilibrium of the single-mode disk accretion model (Eqs. (16) and (18) in Ref. \cite{Klus2014}). In addition, the magnetic field of SXP\,15.3 is measured to be $6\times10^{11}\,$G through the CRSF measurement \citep{Maitra2018}, which is one order of magnitude lower than the estimation by using the single-mode disk accretion model ($6\times10^{12}\,$G; Table\,3 in Ref.\citep{Klus2014}).  In the following, we describe the assumption and formalism of the spin evolution model established in Ref. \citep{Xu2019}.

\subsection{Multi-mode spin evolution model\label{multi-mode-model}}

In most BeXRPs, it has been shown that the angular momentum of the material transferred towards NS is large enough to form an accretion disk around the NS \citep{Klus2014}. We accordingly assume that the X-ray luminosity of BeXRPs is related to disk accretion process. The corresponding accretion rate of XRPs is estimated as 
\begin{equation}
    \dot{M} = \frac{L_{\rm X} R}{GM},
    \label{eq:Lx}
\end{equation}
where we assume that the gravitational potential energy of the accreted material is mainly released at the surface of the NS.

During type I/II outbursts, the accretion rate is above $\sim10^{-10}\,M_\odot\,\mathrm{yr}^{-1}$, suggesting that the accretion disk can be described by the standard thin disk model \citep{Shakura1973}. Different from spherical accretion, accretion disk is expected to slightly penetrate the magnetosphere due to its geometry, and the relation between the inner radius $R_{\rm disk}$ of the disk and the magnetospheric radius  $R_{\rm mag}$ is 
\begin{equation}
    R_{\rm disk} = \phi R_{\rm mag},
\end{equation}
where $\phi$ is about $0.5-1$. The mass infalling at the inner edge transfers the Keplerian orbital angular momentum of the disk material, imposing a spin-up torque $N_0$ on the central NS. Besides the mass infalling, a wide range of accretion disk have angular momentum exchange with the central NS due to the coupling between the magnetic field of pulsar and the disk material. The inner part of the disk transfer angular momentum towards the NS, while the outer part of the disk exerts angular momentum from the pulsar. The total torque $N$ of an accretion disk is measured by a dimensionless torque $n$,
\begin{equation}
    N= N_0 n(\omega),
\end{equation}
where $\omega$ is the fastness parameter, defined as the ratio of the spin angular velocity of NS to the Keplerian angular velocity at the inner edge of accretion disk,
\begin{equation}
    \omega= \Omega_{\rm s}/\Omega_{\rm K}(R_{\rm disk}).
\end{equation}
The form of the dimensionless torque depends on the property of the accretion flow \citep[e.g.,][]{Ghosh1979a,Ghosh1979b,Wang1995}, which is usually approximated by a simplified form \citep{Wang1995} 
\begin{equation}
    n=1-\frac{\omega}{\omega_{\rm crit}}, 
    \label{eq:dimensionless_torque}
\end{equation}
where the critical fastness parameter $\omega_{\rm crit}$ is treated as a constant. 

The quiescent state of BeXRPs is featured by its low X-ray luminosity, which is about three orders of magnitude lower than that of type I outbursts.  The origin of this faint X-ray emission is still unclear. In this model, we assume that it comes from the remaining material from the interaction at periastron passage. Since the corresponding mass accretion rate is much lower than that during type I outbursts, we expect the properties of the accretion flow is more closer to an advection-dominated accretion flow \citep[ADAF;][]{Narayan1995}, where material has considerable radial velocity towards the central object with a sub-Keplerian rotational velocity. The torque of an ADAF imposing on NS is assumed to have a similar form as the thin disk case but corrected by a parameter $A$, capturing sub-Keplerian rotation. The corresponding torque $N_{\rm quiescent}$ is given by
\begin{equation}
    N_{\rm quiescent} = A N_0 n(\omega/A).
\end{equation}
In addition, we also take into account a potential disk wind launched at the radius of the magnetosphere, taking away the Keplerian angular momentum at the radius of the magnetosphere, which imposes an additional spin-down torque on the NS. How strong this torque is depends on the fraction  $\eta$ of mass loaded into this magnetic wind.

The spin evolution is determined by the overall torque $N_{\rm total}$ related to these different accretion modes, which is 
\begin{equation}
    N_{\rm total} = x N_{\rm I} + y N_{\rm II} + (1-x-y)N_{\rm quiescent},
\end{equation}
where $N_{\rm I}$ is the torque related to type I outburst with a time fraction of $x$, $N_{\rm II}$ is the torque related to type II outburst with a time fraction of $y$, and $N_{\rm quiescent}$ is the torque related to quiescent state with a time fraction of $(1-x-y)$.  

We define the (long-term) equilibrium spin period $P_{\rm eq}$ to be the spin period that makes $N_{\rm total}$ equal zero. Ref.\citep{Xu2019} presents the spin evolution of BeXRPs over time with various combinations of input parameters by solving the differential equations of spin evolution of the multi-mode model. These numerical simulations demonstrate that the final average spin period can be well described by the expected $P_{\rm eq}$. Furthermore, the simulations also show that the time $\tau_\text{eq}$ that a BeXRP takes to reach the long-term equilibrium state is below about 1\,Myr. It is longer than the expectation of the single-mode model due to a weaker overall spin-up torque, but still much shorter than the lifetime of Be stars. 
Therefore the observed BeXRPs have probably achieved their long-term equilibrium state. Prior study based on the canonical single-mode model also reaches a similar conclusion by analysing the observed long-term averaged spin period change rates over various luminosity states \citep{Ho2014MNRAS.437.3664H}. Different from the single-mode model, which finally achieves a steady equilibrium period \citep{Ho2014MNRAS.437.3664H}, the multi-mode model expects BeXRPs to show episodic spin-up or spin-down, making the actual spin period of a BeXRP in long-term spin equilibrium oscillate around the expected $P_{\rm eq}$. For example, a BeXRP could show a spin-up mode during an outburst but a spin-down mode during a quiescent phase, and these two modes can achieve a near-zero net torque in long-term evolution, which does not contradict current observations (e.g., Refs.\citep{Klus2014} and \citep{Vasilopoulos2022}). Still, the possibility that the observed BeXRPs are not in spin equilibrium is not excluded theoretically, but we expect most of the BeXRPs in our sample are close to or already reach their long-term equilibrium state.  

Due to the above reasons, we accordingly interpret our observed spin periods as the expected long-term equilibrium periods according to the multi-mode model, which is evaluated by the formula in Ref.\citep{Xu2019}. The following estimation is not significantly affected if a few BeXRPs were far from their spin equilibrium.
In order to investigate the role of type II outbursts, we treat the peak luminosity $L_{\rm II}$ of the type II  outbursts as a free parameter, and we compute the corresponding $P_{\rm eq}$ to obtain the theoretical $L_{\rm II}\text{--}P_{\rm eq}$ correlation. We vary $L_{\rm II}$ from $10^{36}$ to $10^{39}$ ergs$^{-1}$ according to the observed peak luminosities (Figure \ref{fig:bright_corr}), and we extend the luminosity range further up to $10^{42}\,$ergs$^{-1}$ to cover pulsating ultraluminous X-ray
sources (see Sect.\,\ref{sect:pulx}), and we fix the other parameters to reasonable values, which are summarised in Table \ref{tab:inputs-Peq}. Besides the fiducial parameter set in Table \ref{tab:inputs-Peq}, we also explore various parameter combinations in the parameter study in the Supplementary Materials.

As can be seen in Figure \ref{fig:ulx}, the theoretical prediction is broadly consistent with our observation. Different from the traditional single-mode spin evolution model ($L_{\rm X} \propto P_{\rm eq}^{-7/3}$), the adopted multi-mode model successfully reproduces the observed flatter relation between $L_{\rm X}$ and $P_{\rm spin}$ with canonical magnetic fields, $10^{11}\text{--}10^{13}\,\text{G}$, suggested by the current CRSF  measurements of BeXRPs  (e.g., Refs.\cite{Klus2014}, \cite{Jaisawal2016}, and \cite{Maitra2018}, and the $E_{\rm cyc}$ and $B$ columns in Table \ref{tab:log}). Besides the fiducial parameters, the slope of the $L_{\rm X}\text{--}P_{\rm spin}$ relation can also be well reproduced by other parameter combinations with reasonable magnetic fields (see Supplementary Materials). However, we note that an accurate estimations of the magnetic fields of individual BeXRPs in our sample by using the spin-evolution model is not possible at present, due to the uncertainties in key parameters of the spin-evolution theory. While the multi-mode model requires lower magnetic fields to reproduce the observed spin periods (Figure \ref{fig:parameter_study} and Ref.\cite{Xu2019}), the possibility of supercritical magnetic field is not theoretically excluded (see the Low-$\phi$ High-$\omega_{\rm crit}$ case in Figure \ref{fig:parameter_study}), which needs further verification through future accurate magnetic field measurements. In addition, for a detailed magnetic field estimation by using the multi-mode model, we refer to the Markov Chain Monte Carlo simulation in Ref.\cite{Xu2019}, which fit the multi-mode model to the observed BeXRPs in the Small Magellanic Cloud.

\begin{table}[H]
    \centering
    \begin{tabular}{lc}
    \hline
    Parameter & Value\\
     \hline
    NS mass $M_{\rm NS}$  &$1.4~M_{\odot}$ \\
    NS radius $R_{\rm NS}$ & 13\,km \citep{Miller2019}\\
    NS magnetic field $B$ &  $10^{11}$, $10^{12}$,  $10^{13}$\,G\\
     \multicolumn{2}{l}{  [see available CRSF measurements in Table \ref{tab:log}]}\\
    \hline
    critical fastness parameter $\omega_{\rm crit}$ & 0.3\\
    ratio $\phi$ of $R_{\rm disk}/R_{\rm mag}$ & 0.8\\
    fraction of wind mass loss  $\eta$ & 0.2 \\
    \hline
    Type I outburst & \\
    -- peak luminosity $L_{\rm I}$ & $5\times10^{35}\,$erg s$^{-1}$\\
    -- duty cycle $x$ &  0.1\\
    \hline
    Type II outburst & \\
    -- peak luminosity $L_{\rm II}$ & see text\\
    -- duty cycle $y$ & 0.01\\
    \hline
    Quiescence state & \\
    -- luminosity $L_{\rm q}$ &  $10^{33}\,$erg s$^{-1}$\\
    -- duty cycle & $1-x-y$ \\
    -- ADAF parameter $A$ & 0.2 \citep{Narayan1995}\\
\hline
    \end{tabular}
    \caption{Fiducial input parameters of the multi-mode spin evolution model. The theoretical $L_{\rm II}\text{--}P_{\rm eq}$ correlation is presented in Figure \ref{fig:ulx}. For other parameter combinations, we refer to the parameter study presented in the Supplementary Materials (Figure \ref{fig:parameter_study}).}
    \label{tab:inputs-Peq}
\end{table}

\subsection{In comparison with pulsating ultraluminous X-ray sources\label{sect:pulx}}

There are increasing observational evidences that the majority of ultarluminous X-ray sources (ULXs) are high-mass X-ray binaries (HMXBs)  in nearby galaxies \citep{Kaaret2017, King2023}. The empirical luminosity functions are broadly in line with the higher end of the HMXB luminosity function \citep{Swartz2004, Mineo2012}. The optical companions of several ULXs have been identified as young stars \citep{Liu2013, Motch2014}. Dramatically, the detections of coherent pulsations and CRSFs indicate that a significant fraction of ULXs are powered by NSs \citep{Bachetti2014, Brightman2018}. To date, there are six pulsating ultraluminous X-ray sources (PULXs) reported: M82~X-2 \citep{Bachetti2014}, NGC~7793~P13 \citep{Furst2016, Israel2017a},  NGC~5907~ULX1 \citep{Israel2017b}, NGC~300~ULX1 \citep{Carpano2018}, NGC~1313~X-2 \citep{Sathyaprakash2019}, and M51~ULX-7 \citep{Rodriguez2020}. All six PULXs display large flux variations, or even show the hint of transient between the propeller and accretion regimes \citep{Tsygankov2016}. They occupy the same regions with the disk-fed HMXBs in the Corbet diagram, and the super-orbital period $P_{\rm super}$--$P_{\rm orb}$ diagram (Figures 1 and 5 in \citep{Weng2024}). These results suggest that PULXs are analogous to XRPs in the Galaxy, but with higher luminosity. The caution is that PULXs are quite different from the normal XRPs in some aspects, e.g. the X-ray spectra of PULXs are much softer.

\begin{figure}[H]
\centering
\includegraphics[width=0.45\textwidth]{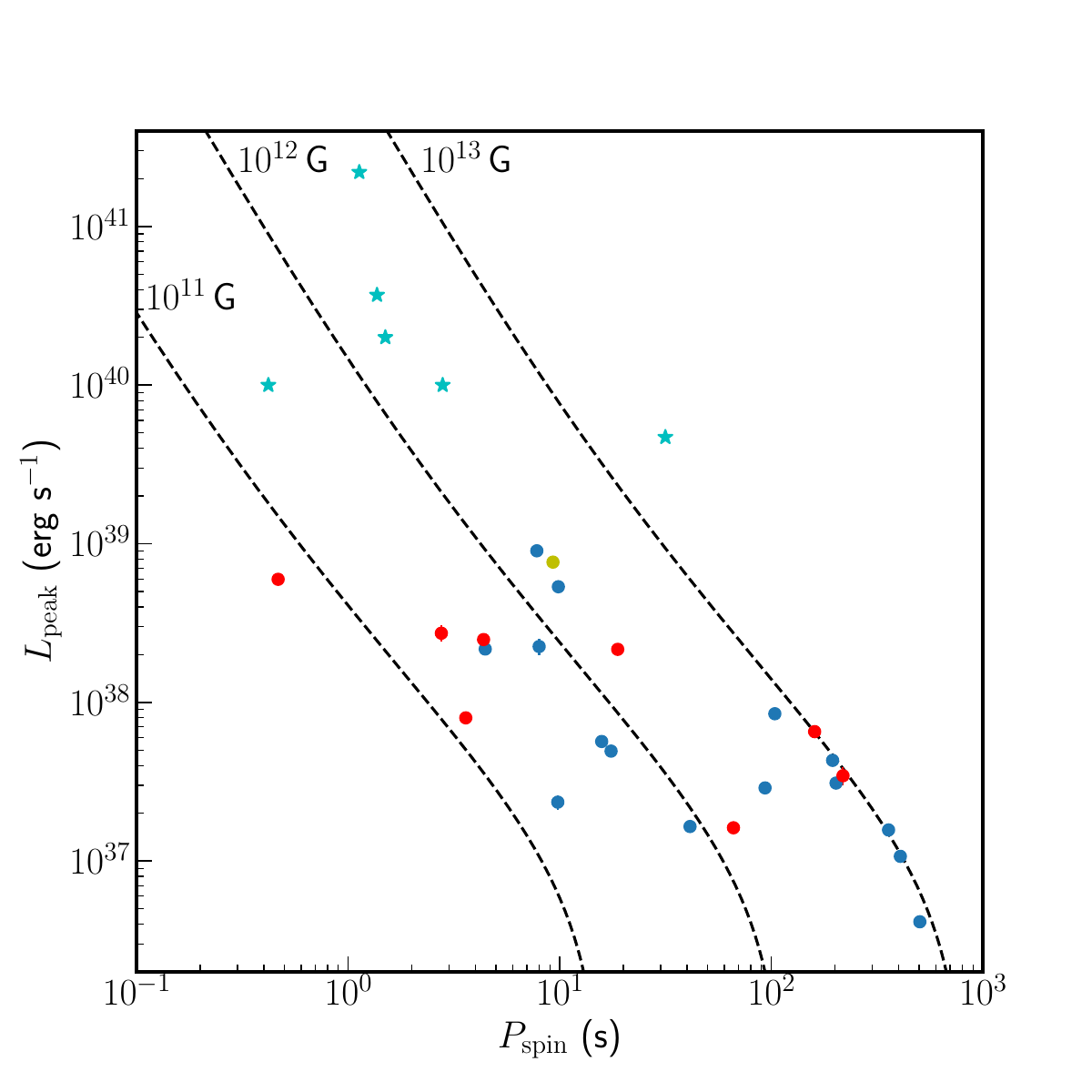}
\caption{$P_{\rm spin}$-$L_{\rm peak}$ diagram for both XRPs and PULXs. Cyan pentacles correspond to PULXs. The dashed lines from lower-left to upper-right correspond to $P_{\rm eq} - L$ relation (more detailed parameters in Section \ref{sec:discussion}) with $B = 10^{11}$, $10^{12}$, $10^{13}$ G indicated by the nearby text of each theoretical curves.
\label{fig:ulx}
}
\end{figure}

Because of their large distances, it is difficult to trace the complete activities of PULXs and to estimate the outburst parameters. We collect the maximum luminosity from the literature (Table \ref{tab:ulx}), and plot the data of PULXs together with the normal XRPs in Figure \ref{fig:ulx}. As can be seen that, the anti-correlation of $P_{\rm spin} - L_{\rm peak}$ becomes even stronger ($\rho/p = -0.821/2.8\times10^{-8}$). The PULXs and the normal XRPs following the same correlation further supports that, both of them are highly magnetized NSs are close to the spin equilibrium determined by the balance of spin-up at the bright state and spin-down during the low-luminosity state. But the outlier, NGC 300 ULX1 has evidently not yet reached the spin equilibrium \citep{Carpano2018, Vasilopoulos2019}.  In this case, we suggest that the X-ray luminosity could be a good $P_{\rm spin}$ indicator. It was suggested that PULXs could spin up to a millisecond period by accreting about 0.1 M$_{\odot}$, becoming powerful sources of gravitational waves \citep{Kluzniak2015}. Because the existence of Eddington limit for highly magnetized NSs \citep{Mushtukov2015}, the correlation should flatten out at the left end (Figure \ref{fig:ulx}).


\begin{table}[H]
    \centering
    \resizebox{0.45\textwidth}{!}{
    \begin{tabular}{cccc}
    \hline
    PULXs & $P_{\rm spin}$ (s) & $L_{\rm peak}$ (erg~s$^{-1}$) & Ref. \\
     \hline
M82~X-2  & 1.37 & $3.7\times10^{40}$ & \cite{Bachetti2014}\\
NGC~7793~P13 & 0.42 & $1\times10^{40}$ & \cite{Furst2016, Israel2017a}\\
NGC~5907~ULX1  & 1.13 & $2.2\times10^{41}$ & \cite{Israel2017b}\\
NGC~300~ULX1 & 31.6 & $4.7\times10^{39}$ & \cite{Carpano2018}\\
NGC~1313~X-2 & 1.5 & $2\times10^{40}$ & \cite{Sathyaprakash2019}\\
M51~ULX-7  & 2.8 & $1\times10^{40}$ & \cite{Rodriguez2020}\\
\hline
    \end{tabular}
    }
    \caption{Parameters of PULXs}
    \label{tab:ulx}
\end{table}


\section{Summary}
\label{sec:sum}

We construct a sample of BeXRPs' giant outbursts by using the {\it RXTE}/ASM, the {\it Swift}/BAT and the {\it NICER} data, estimate the outburst parameters. We find the strong anti-correlations between the spin periods and all four outburst parameters, $L_{\rm peak}$, $T_{\rm 50}$,  $E_{\rm tot}$, and $\bar{L}_{\rm X}$. Taking all 6 PULXs detected in nearby galaxies into account, the anti-correlation between the $L_{\rm peak}$ and the $P_{\rm spin}$ becomes even more significant. However, the sensitivity of the {\it Swift}/BAT and {\it RXTE}/ASM data is too low ($\sim$ 10 mCrab), and these data are unsuitable for studying faint persistent BeXRBs and  extragalactic XRPs. Therefore, we would like to emphasize that the correlations reported in this paper are only applicable to the transient BeXRBs displaying giant outbursts, and caution is required when generalizing them to other populations of BeXRBs.

These results reveal the multi-accretion-mode nature of BeXRPs, and the observed spin periods are likely the outcome  balanced by different accretion modes. This may also be true for other types of accreting NSs with strong X-ray variability, like PULXs. Particularly, the observed $L_{\rm peak}-P_{\rm spin}$ relation also suggests that giant outbursts of BeXRPs greatly contribute to the spin-up of XRPs. XRPs with higher $L_{\rm peak}$ are more likely to accrete more material from donor stars, and thus be spun up to shorter spin periods, and vice versa.

\footnotesize{\textbf{Acknowledgements.}} We thank the anonymous referees for their comments and suggestions, which improved the clarity of the paper. We thank Xi-Yu Chen for valuable discussions. This research has made use of data collected by three NASA's X-ray missions, {\it Swift}, {\it RXTE}, and {\it NICER}, which can be downloaded from the HEASARC database. This work is supported by the China Manned Space Program with grant no CMS-CSST-2025-A13. The authors thank the support from the National Key R\&D Program of China (2021YFA0718500 and 2023YFA1608100) and the National Natural Science Foundation of China (Grants No. 12473041, 12473042, 12393812, 12233002, U2031205, 12041301 and 12121003).

\footnotesize{\textbf{Conflict of interest.}}
The authors declare that they have no conflict of interest.

\bibliographystyle{scpma}

\begin{thebibliography}{10}
\providecommand{\url}[1]{\texttt{#1}}
\providecommand{\urlprefix}{URL }
\providecommand{\doi}[1]{doi:~\href{http://doi.org/#1}{\nolinkurl{#1}}}
\providecommand{\arXiv}[1]{\href{https://arxiv.org/abs/#1}{\nolinkurl{https://arxiv.org/abs/#1}}}
\providecommand{\eprint}[1]{\href{http://arxiv.org/abs/#1}{\nolinkurl{#1}}}

\bibitem{Staubert2019}
R.~{Staubert}, J.~{Tr{\"u}mper}, E.~{Kendziorra}, D.~{Klochkov}, K.~{Postnov},
  P.~{Kretschmar}, K.~{Pottschmidt}, F.~{Haberl}, R.~E. {Rothschild},
  A.~{Santangelo}, J.~{Wilms}, I.~{Kreykenbohm}, and F.~{F{\"u}rst}, \aap
  \textbf{622}, A61 (2019), arXiv: \eprint{1812.03461}.

\bibitem{Kong2022}
L.-D. {Kong}, S.~{Zhang}, S.-N. {Zhang}, L.~{Ji}, V.~{Doroshenko},
  A.~{Santangelo}, Y.-P. {Chen}, F.-J. {Lu}, M.-Y. {Ge}, P.-J. {Wang},
  L.~{Tao}, J.-L. {Qu}, T.-P. {Li}, C.-Z. {Liu}, J.-Y. {Liao}, Z.~{Chang},
  J.-Q. {Peng}, and Q.-C. {Shui}, \apjl \textbf{933}, L3 (2022), arXiv:
  \eprint{2206.04283}.

\bibitem{Liu2005}
Q.~Z. {Liu}, J.~{van Paradijs}, and E.~P.~J. {van den Heuvel}, \aap
  \textbf{442}, 1135 (2005).

\bibitem{Liu2006}
Q.~Z. {Liu}, J.~{van Paradijs}, and E.~P.~J. {van den Heuvel}, \aap
  \textbf{455}, 1165 (2006), arXiv: \eprint{0707.0549}.

\bibitem{Fortin2023}
F.~{Fortin}, F.~{Garc{\'\i}a}, A.~{Simaz Bunzel}, and S.~{Chaty}, \aap
  \textbf{671}, A149 (2023), arXiv: \eprint{2302.02656}.

\bibitem{Neumann2023}
M.~{Neumann}, A.~{Avakyan}, V.~{Doroshenko}, and A.~{Santangelo}, arXiv
  e-prints arXiv:2303.16137 (2023), arXiv: \eprint{2303.16137}.

\bibitem{Reig2011}
P.~{Reig}, \apss \textbf{332}, 1 (2011), arXiv: \eprint{1101.5036}.

\bibitem{Weng2024}
S.-S. {Weng} and L.~{Ji}, Universe \textbf{10}, 453 (2024), arXiv:
  \eprint{2412.17275}.

\bibitem{Okazaki2001}
A.~T. {Okazaki} and I.~{Negueruela}, \aap \textbf{377}, 161 (2001), arXiv:
  \eprint{astro-ph/0108037}.

\bibitem{Okazaki2013}
A.~T. {Okazaki}, K.~{Hayasaki}, and Y.~{Moritani}, \pasj \textbf{65}, 41
  (2013), arXiv: \eprint{1211.5225}.

\bibitem{Finger1996b}
M.~H. {Finger}, R.~B. {Wilson}, and B.~A. {Harmon}, \apj \textbf{459}, 288
  (1996).

\bibitem{Bozzo2009}
E.~{Bozzo}, L.~{Stella}, M.~{Vietri}, and P.~{Ghosh}, \aap \textbf{493}, 809
  (2009), arXiv: \eprint{0811.0049}.

\bibitem{Davies1981}
R.~E. {Davies} and J.~E. {Pringle}, \mnras \textbf{196}, 209 (1981).

\bibitem{Ikhsanov2001}
N.~R. {Ikhsanov}, \aap \textbf{368}, L5 (2001), arXiv:
  \eprint{astro-ph/0111505}.

\bibitem{Shakura2012}
N.~{Shakura}, K.~{Postnov}, A.~{Kochetkova}, and L.~{Hjalmarsdotter}, \mnras
  \textbf{420}, 216 (2012), arXiv: \eprint{1110.3701}.

\bibitem{Weng2017}
S.-S. {Weng}, M.-Y. {Ge}, H.-H. {Zhao}, W.~{Wang}, S.-N. {Zhang}, W.-H. {Bian},
  and Q.-R. {Yuan}, \apj \textbf{843}, 69 (2017), arXiv: \eprint{1701.02983}.

\bibitem{Chashkina2019}
A.~{Chashkina}, G.~{Lipunova}, P.~{Abolmasov}, and J.~{Poutanen}, \aap
  \textbf{626}, A18 (2019), arXiv: \eprint{1902.04609}.

\bibitem{Liu2022}
J.~{Liu}, G.~{Vasilopoulos}, M.~{Ge}, L.~{Ji}, S.-S. {Weng}, S.-N. {Zhang}, and
  X.~{Hou}, \mnras \textbf{517}, 3354 (2022), arXiv: \eprint{2209.11496}.

\bibitem{Klus2014}
H.~{Klus}, W.~C.~G. {Ho}, M.~J. {Coe}, R.~H.~D. {Corbet}, and L.~J. {Townsend},
  \mnras \textbf{437}, 3863 (2014), arXiv: \eprint{1311.4343}.

\bibitem{Xu2019}
X.-T. {Xu} and X.-D. {Li}, \apj \textbf{872}, 102 (2019), arXiv:
  \eprint{1901.04707}.

\bibitem{Krimm2013}
H.~A. {Krimm}, S.~T. {Holland}, R.~H.~D. {Corbet}, A.~B. {Pearlman},
  P.~{Romano}, J.~A. {Kennea}, J.~S. {Bloom}, S.~D. {Barthelmy}, W.~H.
  {Baumgartner}, J.~R. {Cummings}, N.~{Gehrels}, A.~Y. {Lien}, C.~B.
  {Markwardt}, D.~M. {Palmer}, T.~{Sakamoto}, M.~{Stamatikos}, and T.~N.
  {Ukwatta}, \apjs \textbf{209}, 14 (2013), arXiv: \eprint{1309.0755}.

\bibitem{Vasilopoulos2020}
G.~{Vasilopoulos}, P.~S. {Ray}, K.~C. {Gendreau}, P.~A. {Jenke}, G.~K.
  {Jaisawal}, C.~A. {Wilson-Hodge}, T.~E. {Strohmayer}, D.~{Altamirano}, W.~B.
  {Iwakiri}, M.~T. {Wolff}, S.~{Guillot}, C.~{Malacaria}, and A.~L. {Stevens},
  \mnras \textbf{494}, 5350 (2020), arXiv: \eprint{2004.03022}.

\bibitem{DAi2015}
A.~{D'A{\`\i}}, T.~{Di Salvo}, R.~{Iaria}, J.~A. {Garc{\'\i}a}, A.~{Sanna},
  F.~{Pintore}, A.~{Riggio}, L.~{Burderi}, E.~{Bozzo}, T.~{Dauser},
  M.~{Matranga}, C.~G. {Galiano}, and N.~R. {Robba}, \mnras \textbf{449}, 4288
  (2015), arXiv: \eprint{1503.02921}.

\bibitem{Chandra2020}
A.~D. {Chandra}, J.~{Roy}, P.~C. {Agrawal}, and M.~{Choudhury}, \mnras
  \textbf{495}, 2664 (2020), arXiv: \eprint{2004.04930}.

\bibitem{Hou2022}
X.~{Hou}, M.~Y. {Ge}, L.~{Ji}, S.~N. {Zhang}, Y.~{You}, L.~{Tao}, S.~{Zhang},
  R.~{Soria}, H.~{Feng}, M.~{Zhou}, Y.~L. {Tuo}, L.~M. {Song}, and J.~C.
  {Wang}, \apj \textbf{938}, 149 (2022), arXiv: \eprint{2208.14785}.

\bibitem{Wilson2018}
C.~A. {Wilson-Hodge}, C.~{Malacaria}, P.~A. {Jenke}, G.~K. {Jaisawal},
  M.~{Kerr}, M.~T. {Wolff}, Z.~{Arzoumanian}, D.~{Chakrabarty}, J.~P. {Doty},
  K.~C. {Gendreau}, S.~{Guillot}, W.~C.~G. {Ho}, B.~{LaMarr}, C.~B.
  {Markwardt}, F.~{{\"O}zel}, G.~Y. {Prigozhin}, P.~S. {Ray},
  M.~{Ramos-Lerate}, R.~A. {Remillard}, T.~E. {Strohmayer}, M.~L. {Vezie},
  K.~S. {Wood}, and {NICER Science Team}, \apj \textbf{863}, 9 (2018), arXiv:
  \eprint{1806.10094}.

\bibitem{Doroshenko2016}
V.~{Doroshenko}, S.~{Tsygankov}, and A.~{Santangelo}, \aap \textbf{589}, A72
  (2016), arXiv: \eprint{1509.04490}.

\bibitem{Reig1999}
P.~{Reig} and P.~{Roche}, \mnras \textbf{306}, 100 (1999), arXiv:
  \eprint{astro-ph/9902221}.

\bibitem{Kuehnel2014}
M.~{Kuehnel}, M.~H. {Finger}, F.~{Fuerst}, K.~{Pottschmidt}, F.~{Haberl}, and
  J.~{Wilms}, The Astronomer's Telegram \textbf{5856}, 1 (2014).

\bibitem{Yan2012}
J.~{Yan}, J.~A. {Zurita Heras}, S.~{Chaty}, H.~{Li}, and Q.~{Liu}, \apj
  \textbf{753}, 73 (2012), arXiv: \eprint{1205.0063}.

\bibitem{Riquelme2012}
M.~S. {Riquelme}, J.~M. {Torrej{\'o}n}, and I.~{Negueruela}, \aap \textbf{539},
  A114 (2012).

\bibitem{Kuehnel2012}
M.~{Kuehnel}, S.~{Mueller}, I.~{Kreykenbohm}, J.~{Wilms}, K.~{Pottschmidt},
  F.~{Fuerst}, R.~E. {Rothschild}, I.~{Caballero}, D.~{Klochkov},
  R.~{Staubert}, S.~{Suchy}, P.~{Kretschmar}, C.~{Ferrigno}, J.~M. {Torrejon},
  and S.~{Martinez-Nunez}, The Astronomer's Telegram \textbf{4564}, 1 (2012).

\bibitem{Ge2020}
M.~Y. {Ge}, L.~{Ji}, S.~N. {Zhang}, A.~{Santangelo}, C.~Z. {Liu},
  V.~{Doroshenko}, R.~{Staubert}, J.~L. {Qu}, S.~{Zhang}, F.~J. {Lu}, L.~M.
  {Song}, T.~P. {Li}, L.~{Tao}, Y.~P. {Xu}, X.~L. {Cao}, Y.~{Chen}, Q.~C. {Bu},
  C.~{Cai}, Z.~{Chang}, G.~{Chen}, L.~{Chen}, T.~X. {Chen}, Y.~B. {Chen}, Y.~P.
  {Chen}, W.~{Cui}, W.~W. {Cui}, J.~K. {Deng}, Y.~W. {Dong}, Y.~Y. {Du}, M.~X.
  {Fu}, G.~H. {Gao}, H.~{Gao}, M.~{Gao}, Y.~D. {Gu}, J.~{Guan}, C.~C. {Guo},
  D.~W. {Han}, Y.~{Huang}, J.~{Huo}, S.~M. {Jia}, L.~H. {Jiang}, W.~C. {Jiang},
  J.~{Jin}, Y.~J. {Jin}, L.~D. {Kong}, B.~{Li}, C.~K. {Li}, G.~{Li}, M.~S.
  {Li}, W.~{Li}, X.~{Li}, X.~B. {Li}, X.~F. {Li}, Y.~G. {Li}, Z.~W. {Li}, X.~H.
  {Liang}, J.~Y. {Liao}, B.~S. {Liu}, G.~Q. {Liu}, H.~W. {Liu}, X.~J. {Liu},
  Y.~N. {Liu}, B.~{Lu}, X.~F. {Lu}, Q.~{Luo}, T.~{Luo}, X.~{Ma}, B.~{Meng},
  Y.~{Nang}, J.~Y. {Nie}, G.~{Ou}, N.~{Sai}, R.~C. {Shang}, X.~Y. {Song},
  L.~{Sun}, Y.~{Tan}, Y.~L. {Tuo}, C.~{Wang}, G.~F. {Wang}, J.~{Wang}, L.~J.
  {Wang}, W.~S. {Wang}, Y.~D. {Wang}, Y.~S. {Wang}, X.~Y. {Wen}, B.~B. {Wu},
  B.~Y. {Wu}, M.~{Wu}, G.~C. {Xiao}, S.~{Xiao}, S.~L. {Xiong}, H.~{Xu}, J.~W.
  {Yang}, S.~{Yang}, Y.~J. {Yang}, Y.~J. {Yang}, Q.~B. {Yi}, Q.~Q. {Yin},
  Y.~{You}, A.~M. {Zhang}, C.~M. {Zhang}, F.~{Zhang}, H.~M. {Zhang},
  J.~{Zhang}, T.~{Zhang}, W.~C. {Zhang}, W.~{Zhang}, W.~Z. {Zhang}, Y.~{Zhang},
  Y.~F. {Zhang}, Y.~J. {Zhang}, Y.~{Zhang}, Z.~{Zhang}, Z.~{Zhang}, Z.~L.
  {Zhang}, H.~S. {Zhao}, X.~F. {Zhao}, S.~J. {Zheng}, Y.~G. {Zheng}, D.~K.
  {Zhou}, J.~F. {Zhou}, R.~L. {Zhuang}, Y.~X. {Zhu}, and Y.~{Zhu}, \apjl
  \textbf{899}, L19 (2020), arXiv: \eprint{2008.01797}.

\bibitem{Staubert2011}
R.~{Staubert}, K.~{Pottschmidt}, V.~{Doroshenko}, J.~{Wilms}, S.~{Suchy},
  R.~{Rothschild}, and A.~{Santangelo}, \aap \textbf{527}, A7 (2011), arXiv:
  \eprint{1012.2459}.

\bibitem{Orlandini2012}
M.~{Orlandini}, F.~{Frontera}, N.~{Masetti}, V.~{Sguera}, and L.~{Sidoli}, \apj
  \textbf{748}, 86 (2012), arXiv: \eprint{1012.1218}.

\bibitem{Donmez2020}
{\c{C}}.~K. {D{\"o}nmez}, M.~M. {Serim}, S.~{\c{C}}. {{\.I}nam},
  {\c{S}}.~{{\c{S}}ahiner}, D.~{Serim}, and A.~{Baykal}, \mnras \textbf{496},
  1768 (2020), arXiv: \eprint{1911.02871}.

\bibitem{Finger1996a}
M.~H. {Finger}, R.~B. {Wilson}, and D.~{Chakrabarty}, \aaps \textbf{120}, 209
  (1996).

\bibitem{Lutovinov2019}
A.~A. {Lutovinov}, S.~S. {Tsygankov}, D.~I. {Karasev}, S.~V. {Molkov}, and
  V.~{Doroshenko}, \mnras \textbf{485}, 770 (2019), arXiv: \eprint{1902.05153}.

\bibitem{Malacaria2020}
C.~{Malacaria}, P.~{Jenke}, O.~J. {Roberts}, C.~A. {Wilson-Hodge}, W.~H.
  {Cleveland}, B.~{Mailyan}, and {GBM Accreting Pulsars Program Team}, \apj
  \textbf{896}, 90 (2020), arXiv: \eprint{2004.00051}.

\bibitem{Malacaria2023}
C.~{Malacaria}, L.~{Ducci}, M.~{Falanga}, D.~{Altamirano}, E.~{Bozzo},
  S.~{Guillot}, G.~K. {Jaisawal}, P.~{Kretschmar}, M.~{Ng}, P.~{Pradhan},
  R.~{Rothschild}, A.~{Sanna}, P.~{Thalhammer}, and J.~{Wilms}, \aap
  \textbf{669}, A38 (2023), arXiv: \eprint{2211.06367}.

\bibitem{Malacaria2021}
C.~{Malacaria}, P.~{Kretschmar}, K.~K. {Madsen}, C.~A. {Wilson-Hodge}, J.~B.
  {Coley}, P.~{Jenke}, A.~A. {Lutovinov}, K.~{Pottschmidt}, S.~S. {Tsygankov},
  and J.~{Wilms}, \apj \textbf{909}, 153 (2021), arXiv: \eprint{2101.07020}.

\bibitem{Salganik2022}
A.~{Salganik}, S.~S. {Tsygankov}, A.~A. {Djupvik}, D.~I. {Karasev}, A.~A.
  {Lutovinov}, D.~A.~H. {Buckley}, M.~{Gromadzki}, and J.~{Poutanen}, \mnras
  \textbf{509}, 5955 (2022), arXiv: \eprint{2111.08997}.

\bibitem{Tuo2020}
Y.~L. {Tuo}, L.~{Ji}, S.~S. {Tsygankov}, T.~{Mihara}, L.~M. {Song}, M.~Y. {Ge},
  A.~{Nabizadeh}, L.~{Tao}, J.~L. {Qu}, Y.~{Zhang}, S.~{Zhang}, S.~N. {Zhang},
  Q.~C. {Bu}, L.~{Chen}, Y.~P. {Xu}, X.~L. {Cao}, Y.~{Chen}, C.~Z. {Liu},
  C.~{Cai}, Z.~{Chang}, G.~{Chen}, T.~X. {Chen}, Y.~B. {Chen}, Y.~P. {Chen},
  W.~{Cui}, W.~W. {Cui}, J.~K. {Deng}, Y.~W. {Dong}, Y.~Y. {Du}, M.~X. {Fu},
  G.~H. {Gao}, H.~{Gao}, M.~{Gao}, Y.~D. {Gu}, J.~{Guan}, C.~C. {Guo}, D.~W.
  {Han}, Y.~{Huang}, J.~{Huo}, S.~M. {Jia}, L.~H. {Jiang}, W.~C. {Jiang},
  J.~{Jin}, Y.~J. {Jin}, L.~D. {Kong}, B.~{Li}, C.~K. {Li}, G.~{Li}, M.~S.
  {Li}, T.~P. {Li}, W.~{Li}, X.~{Li}, X.~B. {Li}, X.~F. {Li}, Y.~G. {Li}, Z.~W.
  {Li}, X.~H. {Liang}, J.~Y. {Liao}, B.~S. {Liu}, G.~Q. {Liu}, H.~W. {Liu},
  X.~J. {Liu}, Y.~N. {Liu}, B.~{Lu}, F.~J. {Lu}, X.~F. {Lu}, Q.~{Luo},
  T.~{Luo}, X.~{Ma}, B.~{Meng}, Y.~{Nang}, J.~Y. {Nie}, G.~{Ou}, N.~{Sai},
  R.~C. {Shang}, X.~Y. {Song}, L.~{Sun}, Y.~{Tan}, C.~{Wang}, G.~F. {Wang},
  J.~{Wang}, W.~S. {Wang}, Y.~S. {Wang}, X.~Y. {Wen}, B.~Y. {Wu}, B.~B. {Wu},
  M.~{Wu}, G.~C. {Xiao}, S.~{Xiao}, S.~L. {Xiong}, J.~W. {Yang}, S.~{Yang},
  Y.~J. {Yang}, Y.~J. {Yang}, Q.~B. {Yi}, Q.~Q. {Yin}, Y.~{You}, A.~M. {Zhang},
  C.~M. {Zhang}, F.~{Zhang}, H.~M. {Zhang}, J.~{Zhang}, T.~{Zhang}, W.~{Zhang},
  W.~C. {Zhang}, W.~Z. {Zhang}, Y.~{Zhang}, Y.~F. {Zhang}, Y.~J. {Zhang}, Y.~H.
  {Zhang}, Y.~{Zhang}, Z.~{Zhang}, Z.~{Zhang}, Z.~L. {Zhang}, H.~S. {Zhao},
  X.~F. {Zhao}, S.~J. {Zheng}, Y.~G. {Zheng}, D.~K. {Zhou}, J.~F. {Zhou}, Y.~X.
  {Zhu}, Y.~{Zhu}, and R.~L. {Zhuang}, Journal of High Energy Astrophysics
  \textbf{27}, 38 (2020), arXiv: \eprint{2004.13307}.

\bibitem{Beri2021}
A.~{Beri}, T.~{Girdhar}, N.~K. {Iyer}, and C.~{Maitra}, \mnras \textbf{500},
  1350 (2021), arXiv: \eprint{2009.12896}.

\bibitem{Marcu2015}
D.~M. {Marcu-Cheatham}, K.~{Pottschmidt}, M.~{K{\"u}hnel}, S.~{M{\"u}ller},
  S.~{Falkner}, I.~{Caballero}, M.~H. {Finger}, P.~J. {Jenke}, C.~A.
  {Wilson-Hodge}, F.~{F{\"u}rst}, V.~{Grinberg}, P.~B. {Hemphill},
  I.~{Kreykenbohm}, D.~{Klochkov}, R.~E. {Rothschild}, Y.~{Terada}, T.~{Enoto},
  W.~{Iwakiri}, M.~T. {Wolff}, P.~A. {Becker}, K.~S. {Wood}, and J.~{Wilms},
  \apj \textbf{815}, 44 (2015), arXiv: \eprint{1510.05032}.

\bibitem{Doroshenko2017}
V.~{Doroshenko}, S.~S. {Tsygankov}, A.~A. {Mushtukov}, A.~A. {Lutovinov},
  A.~{Santangelo}, V.~F. {Suleimanov}, and J.~{Poutanen}, \mnras \textbf{466},
  2143 (2017), arXiv: \eprint{1607.03933}.

\bibitem{Galloway2004}
D.~K. {Galloway}, E.~H. {Morgan}, and A.~M. {Levine}, \apj \textbf{613}, 1164
  (2004), arXiv: \eprint{astro-ph/0401476}.

\bibitem{Furst2014}
F.~{F{\"u}rst}, K.~{Pottschmidt}, J.~{Wilms}, J.~{Kennea}, M.~{Bachetti},
  E.~{Bellm}, S.~E. {Boggs}, D.~{Chakrabarty}, F.~E. {Christensen}, W.~W.
  {Craig}, C.~J. {Hailey}, F.~{Harrison}, D.~{Stern}, J.~A. {Tomsick}, D.~J.
  {Walton}, and W.~{Zhang}, \apjl \textbf{784}, L40 (2014), arXiv:
  \eprint{1403.1901}.

\bibitem{Wilson2008}
C.~A. {Wilson}, M.~H. {Finger}, and A.~{Camero-Arranz}, \apj \textbf{678}, 1263
  (2008), arXiv: \eprint{0804.1375}.

\bibitem{Yang2024}
W.~{Yang}, W.~{Wang}, and P.~R. {Epili}, \apj \textbf{969}, 107 (2024), arXiv:
  \eprint{2405.03209}.

\bibitem{Wilson1998}
C.~A. {Wilson}, M.~H. {Finger}, B.~A. {Harmon}, D.~{Chakrabarty}, and
  T.~{Strohmayer}, \apj \textbf{499}, 820 (1998), arXiv:
  \eprint{astro-ph/9802324}.

\bibitem{Molkov2019}
S.~{Molkov}, A.~{Lutovinov}, S.~{Tsygankov}, I.~{Mereminskiy}, and
  A.~{Mushtukov}, \apjl \textbf{883}, L11 (2019), arXiv: \eprint{1909.09159}.

\bibitem{Camero2007}
A.~{Camero Arranz}, C.~A. {Wilson}, M.~H. {Finger}, and V.~{Reglero}, \aap
  \textbf{473}, 551 (2007), arXiv: \eprint{astro-ph/0703060}.

\bibitem{Wilson1999}
C.~A. {Wilson}, M.~H. {Finger}, and D.~M. {Scott}, \apj \textbf{511}, 367
  (1999), arXiv: \eprint{astro-ph/9808301}.

\bibitem{Bildsten1997}
L.~{Bildsten} and E.~F. {Brown}, \apj \textbf{477}, 897 (1997), arXiv:
  \eprint{astro-ph/9609155}.

\bibitem{Sanna2017}
A.~{Sanna}, A.~{Riggio}, L.~{Burderi}, F.~{Pintore}, T.~{Di Salvo},
  A.~{D'A{\`\i}}, E.~{Bozzo}, P.~{Esposito}, A.~{Segreto}, F.~{Scarano},
  R.~{Iaria}, and A.~F. {Gambino}, \mnras \textbf{469}, 2 (2017), arXiv:
  \eprint{1703.04449}.

\bibitem{Manikantan2023}
H.~{Manikantan}, B.~{Paul}, and V.~{Rana}, \mnras \textbf{526}, 1 (2023),
  arXiv: \eprint{2308.15129}.

\bibitem{Pietrzynski2019}
G.~{Pietrzy{\'n}ski}, D.~{Graczyk}, A.~{Gallenne}, W.~{Gieren}, I.~B.
  {Thompson}, B.~{Pilecki}, P.~{Karczmarek}, M.~{G{\'o}rski}, K.~{Suchomska},
  M.~{Taormina}, B.~{Zgirski}, P.~{Wielg{\'o}rski}, Z.~{Ko{\l}aczkowski},
  P.~{Konorski}, S.~{Villanova}, N.~{Nardetto}, P.~{Kervella}, F.~{Bresolin},
  R.~P. {Kudritzki}, J.~{Storm}, R.~{Smolec}, and W.~{Narloch}, \nat
  \textbf{567}, 200 (2019), arXiv: \eprint{1903.08096}.

\bibitem{Graczyk2014}
D.~{Graczyk}, G.~{Pietrzy{\'n}ski}, I.~B. {Thompson}, W.~{Gieren},
  B.~{Pilecki}, P.~{Konorski}, A.~{Udalski}, I.~{Soszy{\'n}ski},
  S.~{Villanova}, M.~{G{\'o}rski}, K.~{Suchomska}, P.~{Karczmarek}, R.-P.
  {Kudritzki}, F.~{Bresolin}, and A.~{Gallenne}, \apj \textbf{780}, 59 (2014),
  arXiv: \eprint{1311.2340}.

\bibitem{Kirsch2005}
M.~G. {Kirsch}, U.~G. {Briel}, D.~{Burrows}, S.~{Campana}, G.~{Cusumano},
  K.~{Ebisawa}, M.~J. {Freyberg}, M.~{Guainazzi}, F.~{Haberl}, K.~{Jahoda},
  J.~{Kaastra}, P.~{Kretschmar}, S.~{Larsson}, P.~{Lubi{\'n}ski}, K.~{Mori},
  P.~{Plucinsky}, A.~M. {Pollock}, R.~{Rothschild}, S.~{Sembay}, J.~{Wilms},
  and M.~{Yamamoto}, in \emph{UV, X-Ray, and Gamma-Ray Space Instrumentation
  for Astronomy XIV}, (edited by O.~H.~W. {Siegmund}), volume 5898 of
  \emph{Society of Photo-Optical Instrumentation Engineers (SPIE) Conference
  Series}, 22--33 (2005), arXiv: \eprint{astro-ph/0508235}.

\bibitem{Raguzova2005}
N.~V. {Raguzova} and S.~B. {Popov}, Astronomical and Astrophysical Transactions
  \textbf{24}, 151 (2005), arXiv: \eprint{astro-ph/0505275}.

\bibitem{Sidoli2018}
L.~{Sidoli} and A.~{Paizis}, \mnras \textbf{481}, 2779 (2018), arXiv:
  \eprint{1809.00814}.

\bibitem{Ho2014MNRAS.437.3664H}
W.~C.~G. {Ho}, H.~{Klus}, M.~J. {Coe}, and N.~{Andersson}, \mnras \textbf{437},
  3664 (2014), arXiv: \eprint{1311.1969}.

\bibitem{Xu2025arXiv250323876X}
X.~T. {Xu}, C.~{Sch{\"u}rmann}, N.~{Langer}, C.~{Wang}, A.~{Schootemeijer},
  T.~{Shenar}, A.~{Ercolino}, F.~{Haberl}, B.~{Hastings}, H.~{Jin},
  M.~{Kramer}, D.~{Lennon}, P.~{Marchant}, K.~{Sen}, T.~M. {Tauris}, and S.~E.
  {de Mink}, arXiv e-prints arXiv:2503.23876 (2025), arXiv:
  \eprint{2503.23876}.

\bibitem{Cheng2014}
Z.~Q. {Cheng}, Y.~{Shao}, and X.~D. {Li}, \apj \textbf{786}, 128 (2014), arXiv:
  \eprint{1404.0219}.

\bibitem{Vasilopoulos2022}
G.~{Vasilopoulos}, G.~K. {Jaisawal}, C.~{Maitra}, F.~{Haberl}, P.~{Maggi}, and
  A.~S. {Karaferias}, \aap \textbf{664}, A194 (2022), arXiv:
  \eprint{2206.06396}.

\bibitem{Maitra2018}
C.~{Maitra}, B.~{Paul}, F.~{Haberl}, and G.~{Vasilopoulos}, \mnras
  \textbf{480}, L136 (2018), arXiv: \eprint{1807.10696}.

\bibitem{Shakura1973}
N.~I. {Shakura} and R.~A. {Sunyaev}, \aap \textbf{24}, 337 (1973).

\bibitem{Ghosh1979a}
P.~{Ghosh} and F.~K. {Lamb}, \apj \textbf{234}, 296 (1979).

\bibitem{Ghosh1979b}
P.~{Ghosh} and F.~K. {Lamb}, \apj \textbf{232}, 259 (1979).

\bibitem{Wang1995}
Y.~M. {Wang}, \apjl \textbf{449}, L153 (1995).

\bibitem{Narayan1995}
R.~{Narayan} and I.~{Yi}, \apj \textbf{452}, 710 (1995), arXiv:
  \eprint{astro-ph/9411059}.

\bibitem{Jaisawal2016}
G.~K. {Jaisawal} and S.~{Naik}, \mnras \textbf{461}, L97 (2016), arXiv:
  \eprint{1705.05544}.

\bibitem{Miller2019}
M.~C. {Miller}, F.~K. {Lamb}, A.~J. {Dittmann}, S.~{Bogdanov},
  Z.~{Arzoumanian}, K.~C. {Gendreau}, S.~{Guillot}, A.~K. {Harding}, W.~C.~G.
  {Ho}, J.~M. {Lattimer}, R.~M. {Ludlam}, S.~{Mahmoodifar}, S.~M. {Morsink},
  P.~S. {Ray}, T.~E. {Strohmayer}, K.~S. {Wood}, T.~{Enoto}, R.~{Foster},
  T.~{Okajima}, G.~{Prigozhin}, and Y.~{Soong}, \apjl \textbf{887}, L24 (2019),
  arXiv: \eprint{1912.05705}.

\bibitem{Kaaret2017}
P.~{Kaaret}, H.~{Feng}, and T.~P. {Roberts}, \araa \textbf{55}, 303 (2017),
  arXiv: \eprint{1703.10728}.

\bibitem{King2023}
A.~{King}, J.-P. {Lasota}, and M.~{Middleton}, \nar \textbf{96}, 101672 (2023),
  arXiv: \eprint{2302.10605}.

\bibitem{Swartz2004}
D.~A. {Swartz}, K.~K. {Ghosh}, A.~F. {Tennant}, and K.~{Wu}, \apjs
  \textbf{154}, 519 (2004), arXiv: \eprint{astro-ph/0405498}.

\bibitem{Mineo2012}
S.~{Mineo}, M.~{Gilfanov}, and R.~{Sunyaev}, \mnras \textbf{419}, 2095 (2012),
  arXiv: \eprint{1105.4610}.

\bibitem{Liu2013}
J.-F. {Liu}, J.~N. {Bregman}, Y.~{Bai}, S.~{Justham}, and P.~{Crowther}, \nat
  \textbf{503}, 500 (2013), arXiv: \eprint{1312.0337}.

\bibitem{Motch2014}
C.~{Motch}, M.~W. {Pakull}, R.~{Soria}, F.~{Gris{\'e}}, and
  G.~{Pietrzy{\'n}ski}, \nat \textbf{514}, 198 (2014), arXiv:
  \eprint{1410.4250}.

\bibitem{Bachetti2014}
M.~{Bachetti}, F.~A. {Harrison}, D.~J. {Walton}, B.~W. {Grefenstette},
  D.~{Chakrabarty}, F.~{F{\"u}rst}, D.~{Barret}, A.~{Beloborodov}, S.~E.
  {Boggs}, F.~E. {Christensen}, W.~W. {Craig}, A.~C. {Fabian}, C.~J. {Hailey},
  A.~{Hornschemeier}, V.~{Kaspi}, S.~R. {Kulkarni}, T.~{Maccarone}, J.~M.
  {Miller}, V.~{Rana}, D.~{Stern}, S.~P. {Tendulkar}, J.~{Tomsick}, N.~A.
  {Webb}, and W.~W. {Zhang}, \nat \textbf{514}, 202 (2014), arXiv:
  \eprint{1410.3590}.

\bibitem{Brightman2018}
M.~{Brightman}, F.~A. {Harrison}, F.~{F{\"u}rst}, M.~J. {Middleton}, D.~J.
  {Walton}, D.~{Stern}, A.~C. {Fabian}, M.~{Heida}, D.~{Barret}, and
  M.~{Bachetti}, Nature Astronomy \textbf{2}, 312 (2018), arXiv:
  \eprint{1803.02376}.

\bibitem{Furst2016}
F.~{F{\"u}rst}, D.~J. {Walton}, F.~A. {Harrison}, D.~{Stern}, D.~{Barret},
  M.~{Brightman}, A.~C. {Fabian}, B.~{Grefenstette}, K.~K. {Madsen}, M.~J.
  {Middleton}, J.~M. {Miller}, K.~{Pottschmidt}, A.~{Ptak}, V.~{Rana}, and
  N.~{Webb}, \apjl \textbf{831}, L14 (2016), arXiv: \eprint{1609.07129}.

\bibitem{Israel2017a}
G.~L. {Israel}, A.~{Papitto}, P.~{Esposito}, L.~{Stella}, L.~{Zampieri},
  A.~{Belfiore}, G.~A. {Rodr{\'\i}guez Castillo}, A.~{De Luca}, A.~{Tiengo},
  F.~{Haberl}, J.~{Greiner}, R.~{Salvaterra}, S.~{Sandrelli}, and G.~{Lisini},
  \mnras \textbf{466}, L48 (2017), arXiv: \eprint{1609.06538}.

\bibitem{Israel2017b}
G.~L. {Israel}, A.~{Belfiore}, L.~{Stella}, P.~{Esposito}, P.~{Casella}, A.~{De
  Luca}, M.~{Marelli}, A.~{Papitto}, M.~{Perri}, S.~{Puccetti}, G.~A.~R.
  {Castillo}, D.~{Salvetti}, A.~{Tiengo}, L.~{Zampieri}, D.~{D'Agostino},
  J.~{Greiner}, F.~{Haberl}, G.~{Novara}, R.~{Salvaterra}, R.~{Turolla},
  M.~{Watson}, J.~{Wilms}, and A.~{Wolter}, Science \textbf{355}, 817 (2017),
  arXiv: \eprint{1609.07375}.

\bibitem{Carpano2018}
S.~{Carpano}, F.~{Haberl}, C.~{Maitra}, and G.~{Vasilopoulos}, \mnras
  \textbf{476}, L45 (2018), arXiv: \eprint{1802.10341}.

\bibitem{Sathyaprakash2019}
R.~{Sathyaprakash}, T.~P. {Roberts}, D.~J. {Walton}, F.~{Fuerst},
  M.~{Bachetti}, C.~{Pinto}, W.~N. {Alston}, H.~P. {Earnshaw}, A.~C. {Fabian},
  M.~J. {Middleton}, and R.~{Soria}, \mnras \textbf{488}, L35 (2019), arXiv:
  \eprint{1906.00640}.

\bibitem{Rodriguez2020}
G.~A. {Rodr{\'\i}guez Castillo}, G.~L. {Israel}, A.~{Belfiore},
  F.~{Bernardini}, P.~{Esposito}, F.~{Pintore}, A.~{De Luca}, A.~{Papitto},
  L.~{Stella}, A.~{Tiengo}, L.~{Zampieri}, M.~{Bachetti}, M.~{Brightman},
  P.~{Casella}, D.~{D'Agostino}, S.~{Dall'Osso}, H.~P. {Earnshaw},
  F.~{F{\"u}rst}, F.~{Haberl}, F.~A. {Harrison}, M.~{Mapelli}, M.~{Marelli},
  M.~{Middleton}, C.~{Pinto}, T.~P. {Roberts}, R.~{Salvaterra}, R.~{Turolla},
  D.~J. {Walton}, and A.~{Wolter}, \apj \textbf{895}, 60 (2020), arXiv:
  \eprint{1906.04791}.

\bibitem{Tsygankov2016}
S.~S. {Tsygankov}, A.~A. {Mushtukov}, V.~F. {Suleimanov}, and J.~{Poutanen},
  \mnras \textbf{457}, 1101 (2016), arXiv: \eprint{1507.08288}.

\bibitem{Vasilopoulos2019}
G.~{Vasilopoulos}, M.~{Petropoulou}, F.~{Koliopanos}, P.~S. {Ray}, C.~B.
  {Bailyn}, F.~{Haberl}, and K.~{Gendreau}, \mnras \textbf{488}, 5225 (2019),
  arXiv: \eprint{1905.03740}.

\bibitem{Kluzniak2015}
W.~{Kluzniak} and J.~P. {Lasota}, \mnras \textbf{448}, L43 (2015), arXiv:
  \eprint{1411.1005}.

\bibitem{Mushtukov2015}
A.~A. {Mushtukov}, V.~F. {Suleimanov}, S.~S. {Tsygankov}, and J.~{Poutanen},
  \mnras \textbf{447}, 1847 (2015), arXiv: \eprint{1409.6457}.

\bibitem{Corbet1986}
R.~H.~D. {Corbet}, \mnras \textbf{220}, 1047 (1986).

\end{thebibliography}

\clearpage
\begin{appendix}

\renewcommand{\thefigure}{A\arabic{figure}}
\renewcommand{\thetable}{A\arabic{table}}

\renewcommand{\thesection}{Supplementary Materials\\}
\section{Giant outbursts of XRPs}

\begin{figure*}
\centering
\includegraphics[width=1.0\textwidth]{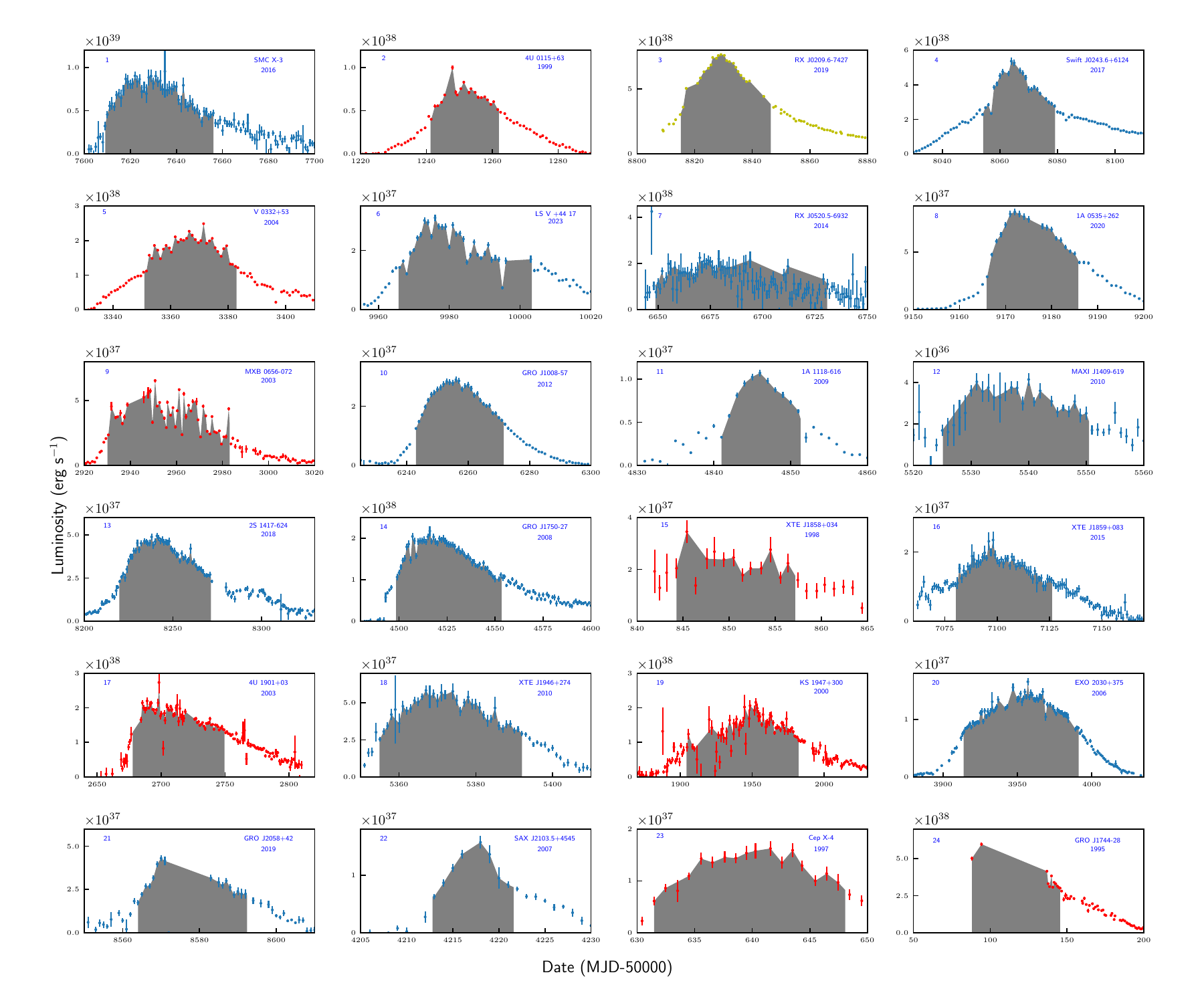}
\caption{{\textbf{Light curves of giant outbursts.} For each source, only the brightest outbursts are displayed here. The light curves from {\it Swift}/BAT and {\it RXTE}/ASM are daily averaged, and are shown with the blue and red crosses, respectively. The data of RX~J0209.6-7427 from the {\it NICER} database, shown in yellow, are averaged for each individual observation. The data points with the luminosity larger than the half of peak luminosity ($L > 0.5 \times L_{\rm peak}$) are adopted to enclose the gray regions and to calculate $E_{\rm tot}$. 
}
\label{fig:bright} 
}
\end{figure*}

\begin{figure*} 
\centering
\includegraphics[width=1.0\textwidth]{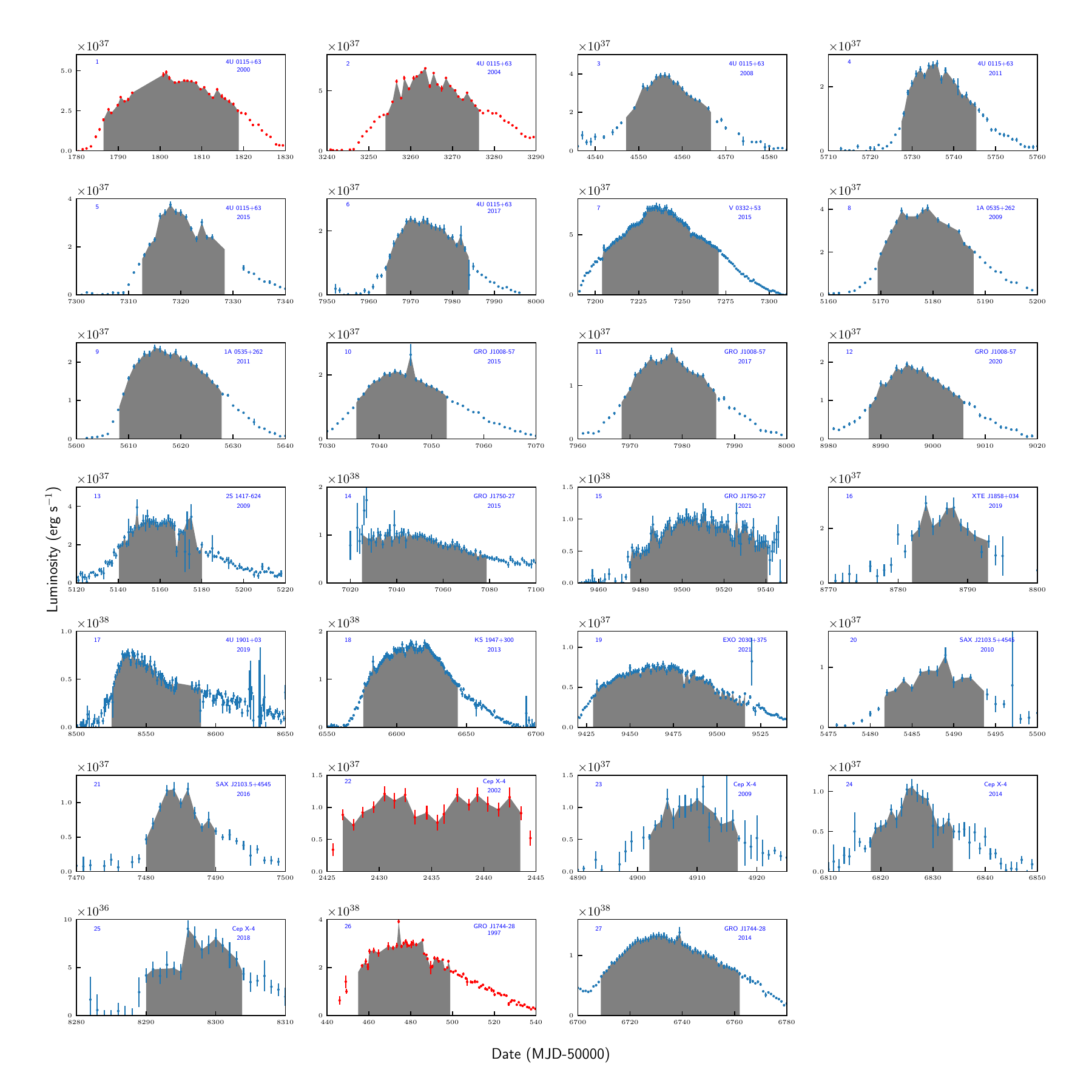}
\caption{{Same as Figure \ref{fig:bright} but for all other less energetic giant outbursts. }
\label{fig:faint} }
\end{figure*}

\begin{figure*} 
\centering
\includegraphics[width=0.9\textwidth]{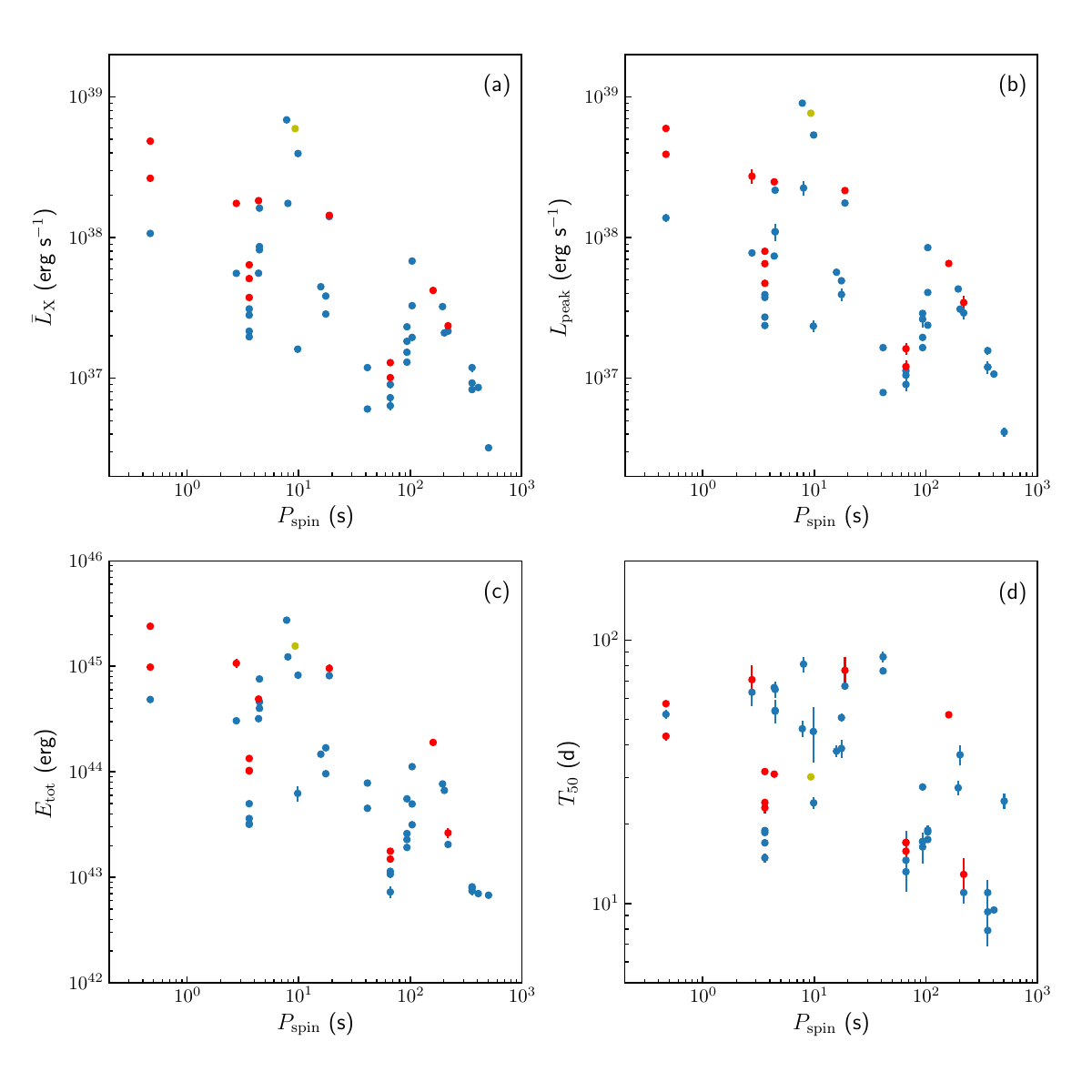}
\caption{{Same as Figure \ref{fig:bright_corr} but for all 51 giant outbursts.}
\label{fig:all_corr} }
\end{figure*}

With the {\it RXTE}/ASM, the {\it Swift}/BAT and {\it NICER} database, we provide the outburst parameters of 51 giant outbursts from 23 BeXRPs and a low-mass X-ray binary, GRO~J1744-28. The light curves of the brightest outburst for each source are displayed in Figure \ref{fig:bright},  and the other outbursts are presented in Figure \ref{fig:faint}. The correlations between the spin periods and four outburst parameters for all 51 outbursts are shown in Figure \ref{fig:all_corr}. The other non-significant correlations between each pair of parameters can be found in Figures \ref{fig:bright_zero} and \ref{fig:all_zero}.

Although most of the BeXRPs occupy a narrow strip in the spin period versus orbital period diagram (i.e. the Corbet diagram \citep{Corbet1986}; Figure \ref{fig:corbet}), the positive correlation is blurred by three XRPs (1A~1118-616, MAXI~J1409-619, and SAX~J2103.5+4545) with long spin periods  ($P_{\rm spin} > 300$ s) and narrow orbits ($P_{\rm orb} < 25$ days). A complete, updated  version of Corbet diagram can be found in Figure 1 of \citep{Weng2024}.

\begin{figure*} 
\centering
\includegraphics[width=1.0\textwidth]{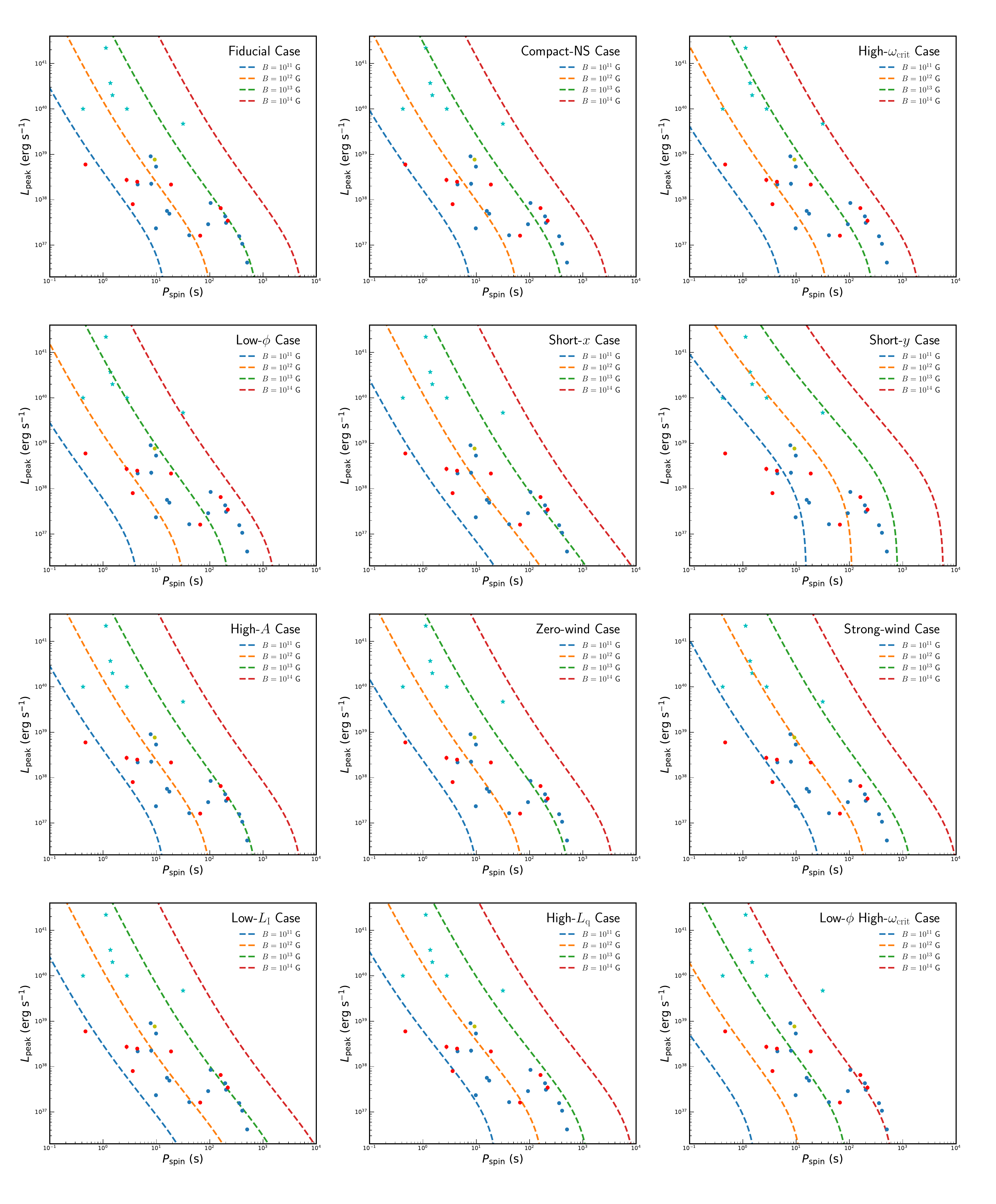}
\caption{Parameter study of the multi-mode spin evolution model. The definition of each cases is listed in Table \ref{tab:parameter-study}. The X-axis is the spin period of BeXRPs, and the Y-axis is the X-ray luminosity of type II outbursts. The dots are observed systems, which have the same meaning as Figure \ref{fig:ulx}. Four dashed lines are the $L_{\rm II}\text{-}P_{\rm eq}$ relation expected by the multi-mode spin evolution model, with four different magnetic fields $10^{11}\,$G (blue), $10^{12}\,$ G (orange), $10^{13}\,$G (green), and $10^{14}\,$G (red). }
  \label{fig:parameter_study}
\end{figure*}

\begin{figure*} 
\centering
\includegraphics[width=\textwidth]{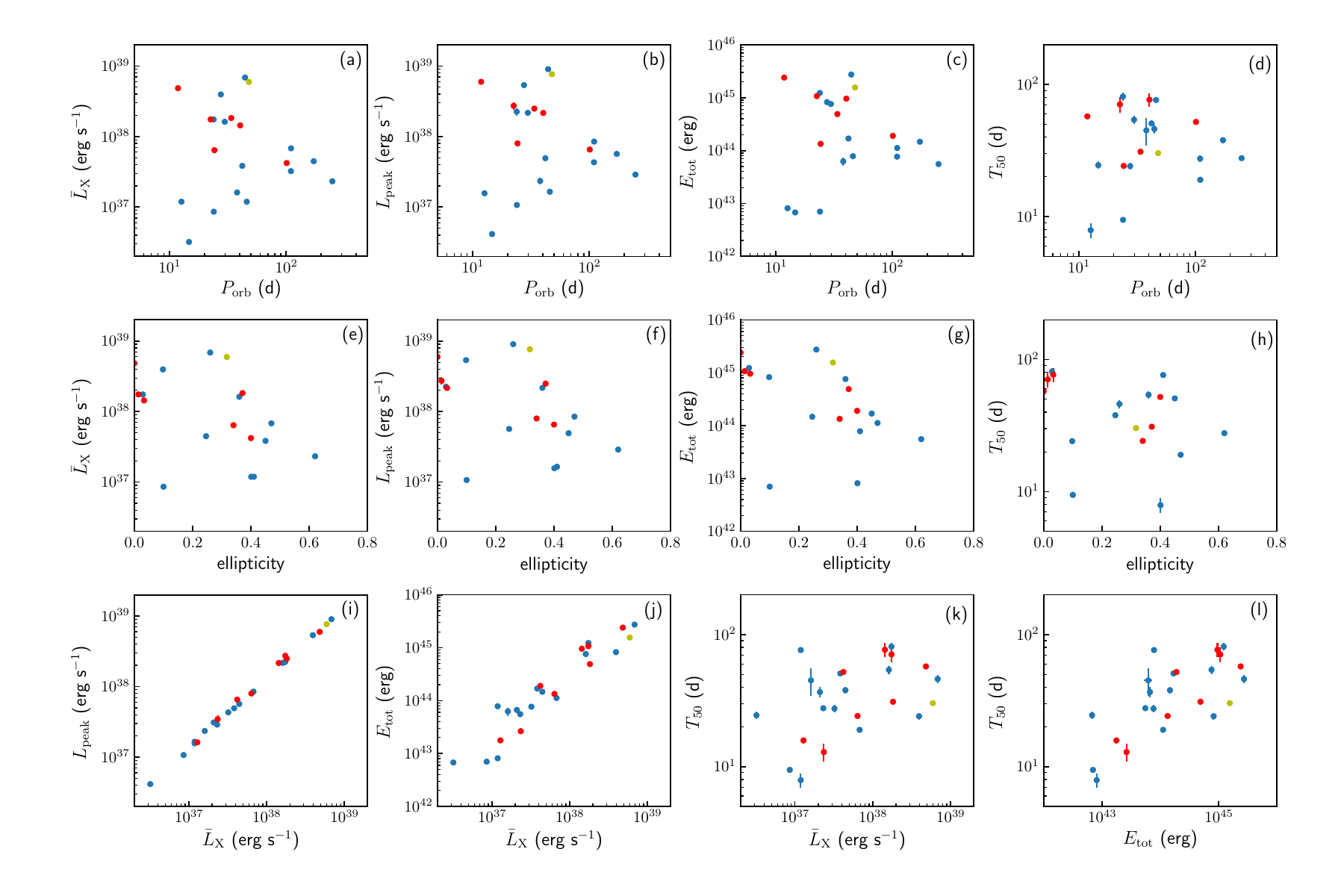}
\caption{\textbf{Correlations between the outburst parameters and the binary parameters.} For each source, only the brightest outbursts are displayed here. panels a-d: $P_{\rm orb}$ versus the outburst parameters; panels e-h: orbital eccentricity versus the outburst parameters; panels i-l: correlations among the outburst parameters.
\label{fig:bright_zero}}
\end{figure*}

\begin{figure*} 
\centering
\includegraphics[width=1.0\textwidth]{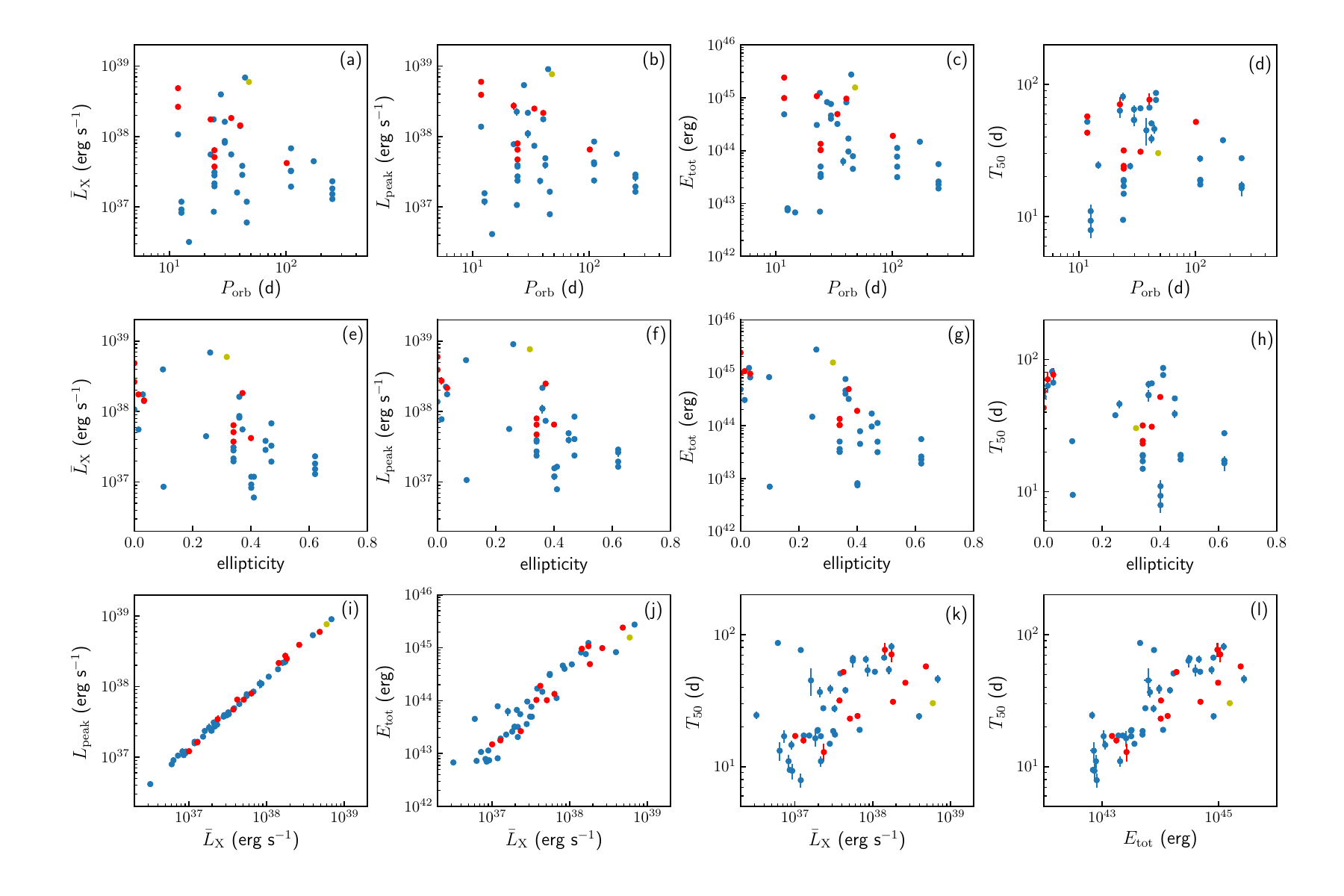}
\caption{Same as Figure \ref{fig:bright_zero} but for all 51 giant outbursts.
\label{fig:all_zero} }
\end{figure*}

\section{Parameter study of the multi-mode spin evolution model}

In the main text, we have adopted the multi-mode spin evolution model established by Ref. \citep{Xu2019} to understand our observed $L_\text{peak}-P_\text{spin}$ correlation of BeXRPs. Different from the canonical single-mode spin evolution model (e.g., Refs.\citep{Klus2014} and \citep{Ho2014MNRAS.437.3664H}, and references therein), the multi-mode model suggests that different X-ray luminosity states correspond to different accretion modes, and needs to be treated separately, which we think is suitable for accreting neutron stars showing strong X-ray variability, like BeXRPs. In this section, we provide a parameter study of the multi-mode spin evolution model to illustrate the uncertainties in evaluating the expected equilibrium spin period $P_{\rm eq}$. Similar to Section \ref{sec:discussion}, we treat the luminosity of the type II outburst as a free parameter, and compute the corresponding $P_{\rm eq}$. The considered cases and the corresponding parameter combinations are listed in Table \ref{tab:parameter-study}. Our results are shown in Figure \ref{fig:parameter_study}. Different from the main text, we compute with four magnetic fields,  $10^{11}$, $10^{12}$, $10^{13}$, and $10^{14}$\,G. The parameters adopted in the main text is listed as the Fiducial case for reference.

\newcommand{\Peqmax}{P_{\rm eq,max}(10^{14}\,\text{G})}

\begin{itemize}
    \item \textbf{Compact-NS case.} We consider a more compact neutron star in this case by setting the neutron star radius to 10 km (13 km in the Fiducial case), which results in a slightly shorter $P_{\rm eq}$ compared to the Fiducial case. The predicted longest spin period for $B=10^{14}\,$G (hereafter $P_{\rm eq,max}(10^{14}\,\text{G})$) shifts from about 5$\times10^3\,$s to $3\times10^3\,$s. The reason is that a shorter radius leads to a lower magnetic moment, $BR_{\rm NS}^3$. 
    \item \textbf{High-$\omega_{\rm crit}$ case.} With $\omega_{\rm crit}=0.8$, we obtain $\Peqmax\simeq2\times10^3\,$s. According to the simplified form of the dimensionless torque (Eq.\,\ref{eq:dimensionless_torque}), a higher $\omega_{\rm crit}$ results in a stronger spin-up torque. As a consequence, the expected $P_{\rm eq}$ is generally lower than the Fiducial case.
    \item \textbf{Low-$\phi$ case.} It is also unclear how far an accretion disk can penetrate the magnetosphere of the neutron star (e.g., Ref.\citep{Sanna2017}). If the radius of the inner edge of the accretion becomes smaller, the region of the accretion disk that transfers angular momentum to the central neutron star (i.e., the region within the co-rotation radius) becomes larger, which enhances the spin-up component in the model. Consequently, we obtain shorter $P_{\rm eq}$, with $\Peqmax\simeq1.5\times10^3\,s$.
    \item \textbf{Short-$x$ case.} By reducing $x$ to 0.01, we find that the choice of the duty cycle of the type I outbursts play a minor role in determining the $L_{\rm peak}\text{-}P_{\rm spin}$ relation. It becomes considerable when the peak luminosity of the type II outbursts becomes comparable to the luminosity of type I outbursts.
    \item \textbf{Short-$y$ case.} We reduce the duty cycle $y$ of type II outbursts to 0.001, which significantly alter the shape of the expected $L_{\rm peak}\text{-}P_{\rm spin}$ correlation. Also, the requirement on magnetic fields for reproducing observations becomes much lower. Those BeXRPs with $P_{\rm spin}$ above 100\,s can be reproduced with magnetic fields close to $10^{11}\,$G. As we have concluded in the main text, type II outbursts greatly contribute to the spin-up of BeXRPs. For a weaker type II outburst, the expected $P_{\rm eq}$ becomes much longer, with $\Peqmax\simeq6\times10^{3}\,$s in the Short-$y$ case.
    \item \textbf{High-$A$ case.} In the case of $A=1$, the accretion flow in the quiescent phase restores the thin accretion disk geometry that the disk material rotates at Keplerian orbital velocity, which may lead to a stronger spin-down torque in the quiescent phase.  In this study, we find that the change in the expected $L_{\rm peak}\text{-}P_{\rm spin}$ correlation is ignorable, as the quiescent phase is too faint compared to the type II outburst.
    \item \textbf{Zero-wind and Strong-wind cases.} The authors of ref.\citep{Xu2019} takes into account the possible mass ejection at the magnetosphere of the neutron star, which is referred to as disk wind in Ref.\citep{Xu2019}, causing a spin-down torque. In the main text, we have adopted a mild wind parameter $\eta=0.2$. In the Zero-wind case, we turn off disk wind by setting $\eta=0$. As expected, the theoretical $P_{\rm eq}$ is shifted towards short-$P_{\rm spin}$ regime, with $\Peqmax\simeq 2\times10^{3}\,$s. On the contrary to the Zero-wind case, the Strong-wind case ($\eta=0.5$) gives a much longer $P_{\rm eq}$ as expected, with $\Peqmax\simeq 10^{4}\,$s.
    \item \textbf{Low-$L_{\rm I}$ and High-$L_{\rm q}$ cases.} In these two cases, we experiment with different luminosities for the type I outburst and quiescent phase. In the Low-$L_{\rm I}$ case, we reduce $L_{\rm I}$ by one order of magnitude, from $5\times10^{35}$\,erg\,s$^{-1}$ to $5\times10^{34}$\,erg\,s$^{-1}$, which slightly reduce the spin-up component in the model. The expected $L_{\rm peak}\text{-}P_{\rm spin}$ correlation is not  significantly affected, and the choice of $L_{\rm I}$ is considerable only if the peak luminosity of the type II outburst becomes comparable to that of the type I outburst. Similarly, increasing $L_{\rm q}$ strengthens the spin-down component in the model, leading to longer $P_{\rm eq}$.
    \item \textbf{Low-$\phi$ High-$\omega_{\rm crit}$ case.} As we discussed in the Low-$\phi$ and High-$\omega_{\rm crit}$ cases, decreasing $\phi$ or increasing $\omega_{\rm crit}$ strengthen the spin-up component in the multi-mode spin evolution model. In this case, we combine the Low-$\phi$ case and High-$\omega_{\rm crit}$ case to explore the shortest $P_{\rm eq}$ that can be obtained with reasonable parameters. Our results suggest that most of the BeXRPs with $P_{\rm spin}$ above 100\,s is consistent with a magnetic field of 10$^{14}\,G$ with low $\phi$ and high $\omega_{\rm crit}$.
\end{itemize}

Our parameter study confirms our conclusion in the main text that the type II outburst plays an dominating role in the spin-up of BeXRPs. Considering the scatter in magnetic fields, we conclude that the observed flatter $L_{\rm peak}\text{--}P_{\rm spin}$ correlation is solidly reproduced by the multi-mode spin evolution model regardless of the choice of input parameters. However, our calculations also suggest that an accurate estimations of magnetic fields of the BeXRPs in our sample is not possible at present, due to the uncertainties in key parameters of spin-evolution theory, like $\omega_{\rm crit}$, $\phi$, $y$, and $\eta$. Generally, the multi-mode spin evolution model requires a weaker magnetic field to reproduce the spin period of the observed BeXRPs than the single-mode model (e.g., \citep{Klus2014}), which is consistent with currently available CRSF measurements (the $E_{\rm cyc}$ column in Table \ref{tab:log} of the main text).

\begin{figure} [H]
    \centering
    \includegraphics[width=0.45\textwidth]{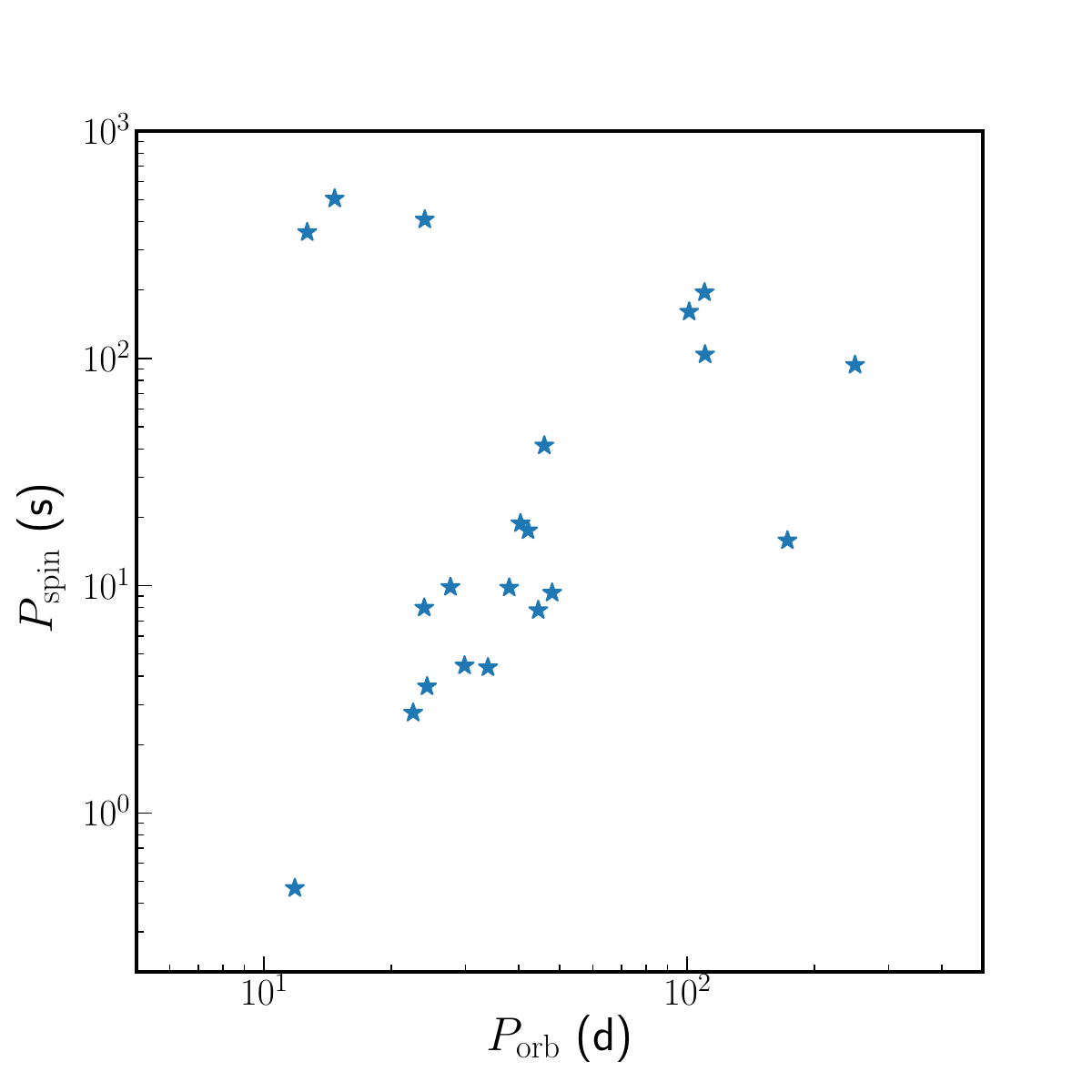}
    \caption{{\textbf{$P_{\rm orb}$-$P_{\rm spin}$ diagram.}}  \label{fig:corbet}}
    \end{figure}

\begin{table*}[!htbp]
    \centering
    \begin{tabular}{lccccccccccc}
    \hline
    & $R_{\rm NS}$ & $\omega_{\rm crit}$ &  $\phi$ &$x$ & $y$ &  $A$ & $\eta$ & $L_{\rm I}$ & $L_{\rm q}$ \\
    & [km] &  &  & &  & & &  [erg s$^{-1}$] & [erg s$^{-1}$] \\
     \hline
      Fiducial & 13 & 0.3 & 0.8 & 0.1 & 0.01 & 0.2 & 0.2 & $5\times10^{35}$&  $10^{33}$\\
      Compact-NS & 10 &=&=&=&=&=&=&=&=\\
      High-$\omega_{\rm crit}$ &=& 0.8 &=&=&=&=&=&=&=\\
      Low-$\phi$ &=&=& 0.3 &=&=&=&=&=&=\\
       Short-$x$ &=&=&=&0.01&=&=&=&=&=\\
       Short-$y$ &=&=&=&=&0.001&=&=&=&= \\
       High-$A$ &=&=&=&=&=&1&=&=&= \\
       Zero-wind &=&=&=&=&=&=&0&=&=\\
       Strong-wind &=&=&=&=&=&=&0.5&=&=\\
       Low-$L_{\rm I}$ &=&=&=&=&=&=&=&$5\times10^{34}$ &=\\
       High-$L_{\rm q}$ &=&=&=&=&=&=&=&=&$10^{34}$ \\
       Low-$\phi$ High-$\omega_{\rm crit}$ &=&0.8&0.3&=&=&=&=&=&= \\
\hline
    \end{tabular}
    \caption{Input parameters of the cases considered for the parameter study of the multi-mode spin evolution model. We refer to Table \ref{tab:inputs-Peq} for the definitions of the parameters. The parameters adopted in the main text are listed as the Fiducial case for reference. In this table, "=" means that the parameter has the same values as the Fiducial case.  The luminosity of type II outbursts is treated as a free parameter (see Section \ref{sec:discussion}).}
    \label{tab:parameter-study}
\end{table*}

\end{appendix}

\end{multicols}

\end{document}